\documentclass[journal=langd5,manuscript=article,layout=twocolumn,sections=true]{achemso}
\usepackage{natbib}

\usepackage{comment}

\usepackage{booktabs} %
\usepackage{tabularx} %
\usepackage{makecell} %
\usepackage{amsmath}  %
\usepackage[version=4]{mhchem} %
\usepackage{siunitx}  %

\usepackage{chemformula} %
\usepackage[T1]{fontenc} %
\SectionNumbersOn
\author{Muhammad Haroon}
\email{mxh1102@case.edu}
\author{Mitchell Lau}

\author{Christopher L. Wirth}
\email{clw22@case.edu}
\affiliation[Case Western Reserve University]
{Department of Chemical and Biomolecular Engineering, Case Western Reserve University, Cleveland, OH 44106}

\title[Reproducible Measurements in Active Janus Particle Systems]
  {
  Impact of Surface Preparation on the Propulsion of Catalytic Janus Colloids and Implications for Reproducibility
 }

\keywords{Active colloids, Janus particles, Reproducible measurements, Nano-motors}
\begin{document}

\onecolumn

\begin{abstract}
Platinum-coated polystyrene (PS) Janus particles (JPs) propelled by the catalytic decomposition of hydrogen peroxide (H$_2$O$_2$) are a widely used model system for active matter, yet reported speeds for similar particles and fuel concentrations differ by more than an order of magnitude. Broad adoption of these materials requires increasing consistency in the preparation and achieving improved repeatability in their behavior. Herein, we focused on two routine, but rarely reported, preparation steps that we found to alter the particle speed. First, we found that pre-exposing the platinum cap to H$_2$O$_2$ before cell assembly increased the mean speed by a factor of 2.4. The size of this effect depended on batch history, where an aged batch responded strongly, while a freshly prepared batch sped up by only 16\%. Second, we found that plasma cleaning of the glass substrate decreased the speed 2.7 times. In a 120~$\mu$m cell the suppression was the same at both walls even when only one wall was treated. The suppression grew with the treated glass area per unit fluid volume, which pointed to species released into the bulk solution rather than to a change at the wall. Plasma-cleaned glass released more ions into water than untreated glass, and XPS detected an ionic metal fluoride on the plasma-cleaned surface. A model in which the released ions lower the speed through the solution conductivity reproduced the measured speeds in all four cell geometries. Of the 17 published studies compared here, pre-activation is reported in only four studies and substrate cleaning in only six studies. We propose a minimum reporting standard for the preparation and associated speed measurements of catalytic Janus particles.
\end{abstract}

\section{Introduction}
Active matter is broadly defined as systems that transduce energy from their surroundings into directed mechanical motion\cite{ramaswamy_active_2017}. It spans an enormous range of length scales, from molecular motors at the nanometer scale and swimming bacteria at the micrometer scale to fish, birds and larger organisms\cite{cichos_machine_2020}. At the microscale, motion occurs at low Reynolds numbers ($Re \approx 10^{-5}$ for a 3~$\mu$m particle moving at a few $\mu$m/s), where inertia is negligible and viscous drag dominates. Propulsion is coupled to thermal fluctuations in both position and orientation, and this coupling produces non-equilibrium behavior with no macroscale analog. Synthetic micro-swimmers have been proposed for targeted drug delivery\cite{xu_sperm-hybrid_2018}, cargo transport\cite{sundararajan_catalytic_2008}, minimally invasive surgery and bio-sensing\cite{wang_nanomicroscale_2012}, water purification\cite{soler_self-propelled_2013}, oil recovery\cite{tohidi_application_2022}, crack repair\cite{li_self-propelled_2015} and environmental remediation\cite{shields_evolution_2017}, and they serve as model systems for biological propulsion\cite{nsamela_colloidal_2023}. Active colloids can also be driven by external electric or magnetic fields\cite{gauri2026surface,patel2026field}.

As is the case for all microscale phenomena, interfaces play a critical role for both biological and synthetic swimmers. Microorganisms accumulate near surfaces such as vegetation and the ocean floor\cite{ramia_role_1993}, and boundary proximity changes the speed, orientation and trajectory of active particles through hydrodynamic, electrostatic and chemical interactions\cite{simmchen_topographical_2016,uspal_self-propulsion_2014,rashidi_motion_2017,das_boundaries_2015,mozaffari_self-diffusiophoretic_2016}. Understanding these boundary effects is therefore central to both the fundamental science and the use of microscale active matter.

Platinum-coated polystyrene (PS) Janus particles (JPs) are among the most widely used synthetic micro-swimmers. First demonstrated by Howse et al., they consist of a micrometer-scale PS sphere with one hemisphere coated by a thin platinum film, and they self-propel in aqueous hydrogen peroxide (H$_2$O$_2$)\cite{howse_self-motile_2007}. Asymmetric decomposition of peroxide into water and oxygen drives the particle toward its polystyrene side\cite{howse_self-motile_2007}. Speed increases with fuel concentration and decreases with particle size, with typical values of 0.5--10~$\mu$m/s in 1--10\% H$_2$O$_2$ for 1--5~$\mu$m particles\cite{howse_self-motile_2007, ebbens_size_2012, das_floor-_2020}. Platinum films of 5--20~nm are common, and the coverage is usually non-uniform, forming a crescent\cite{das_floor-_2020,rashidi_local_2018,issa_engineered_2023}. The propulsion mechanism is still debated, with ionic self-diffusiophoresis and self-electrophoresis both implicated\cite{brown_ionic_2014, wang_open_2023,bishop2023active}. Near a substrate, particles tend to slide along the wall, a state set by hydrodynamic interactions, the gravitational torque of the cap and orientational Brownian motion\cite{das_boundaries_2015, simmchen_topographical_2016, das_floor-_2020}. Platinum is much denser than polystyrene (21.45 versus 1.05~g/cm$^3$), and with a thick enough cap the gravitational torque can suppress rotational diffusion and propulsion\cite{rashidi_influence_2020, campbell_gravitaxis_2013}. A recent study showed that thicker caps reduce speed and quench orientation\cite{haroon_coupled_2026}. Boundary chemistry matters as well: particles have been reported to move faster near hydrophobic than near hydrophilic surfaces\cite{ketzetzi_slip_2020}, and ions released from treated substrates can reduce speed throughout a cell, including at the untreated opposite wall\cite{haroon_coupled_2026}. This sensitivity makes Pt/PS JPs a controllable platform, but it also means that small differences in experimental procedure can produce large differences in measured behavior.

\begin{table*}[htbp]
  \setlength{\tabcolsep}{3pt}
  \footnotesize
  \caption{Experimental parameters reported for catalytic Janus-sphere swimmers}
  \label{tbl:literature_review}
  \begin{tabularx}{\textwidth}{
    l
    >{\raggedright\arraybackslash}X
    c
    c
    c
    >{\raggedright\arraybackslash}X
    c
  }
    \toprule
    \textbf{Author} & \textbf{Material/Size (D)} & \textbf{Pt (nm)} & \textbf{[H$_2$O$_2$]} & \textbf{Speed ($\mu$m/s)} & \textbf{Substrate} & \textbf{Pre-act.} \\
    \midrule
    Ebbens \textit{et al.}\cite{ebbens_size_2012} & 0.5--10 $\mu$m & NR & 10\% & 2--50 & NR & NR \\
    Raman \textit{et al.}\cite{raman_dynamics_2023} & SiO$_2$ ($\sim$5 $\mu$m), PS ($\sim$2 $\mu$m) & 15 & 3 wt\% & 0.1--5 & Glass (IPA sonicated, O$_2$ plasma) & NR \\
    Meng \textit{et al.}\cite{meng_effect_2022} & PS (1--5 $\mu$m) & 1--10 & 3\% & 1--7 & ITO glass slides & NR \\
    Baghdady \textit{et al.}\cite{baghdady_enhancing_2024} & PS (1 $\mu$m) & 5 & 6\% & 10--60 & PEG-modified coverslips & NR \\
    Raj \textit{et al.}\cite{raj_direct_2025} & 3 $\mu$m & 125--135 & 5\% w/v & 1--5 & Glass slide \& coverslip & Yes$^a$ \\
    Campbell \textit{et al.}\cite{campbell_gravitaxis_2013} & 1.9, 3.1, 4.8 $\mu$m & 10 & 10 wt\% & 1.5--4 & O$_2$ plasma cleaned cuvette & Yes$^b$ \\
    Howse \textit{et al.}\cite{howse_self-motile_2007} & 1.62 $\mu$m & 5 & 0--10 vol\% & 0--3.5 & Piranha-cleaned cuvette & NR \\
    Brown \textit{et al.}\cite{brown_ionic_2014} & PS (2 $\mu$m) & 5 & 10\% v/v & $\sim$4--20 & Glass capillaries & NR \\
    Simmchen \textit{et al.}\cite{simmchen_topographical_2016} & 2 or 5 $\mu$m & 7 & 5 v/v\% & NR & O$_2$ plasma patterned & NR \\
    Das \textit{et al.}\cite{das_floor-_2020} & PS (2.4, 5 $\mu$m), SiO$_2$ (1, 4 $\mu$m) & 9 & 3 v/v\% & 1.5--2.1 & Glass slide/coverslip & NR \\
    Issa \textit{et al.}\cite{issa_charged_2019} & 5 $\mu$m & 20 & 0--10\% & 0--1.5 & NR & NR \\
    Lyu \textit{et al.}\cite{lyu_active_2021} & SiO$_2$ (5 $\mu$m) & 10 & 5 wt\% & 1--5 & Glass slide & NR \\
    Issa \textit{et al.}\cite{issa_engineered_2023} & PS (3 $\mu$m) & 20 & 3\% & 0.2--1.4 & Glass rinsed w/ IPA \& water & NR \\
    Liu \textit{et al.}\cite{liu_quantifying_2025} & PS (5 $\mu$m) & 15 nm(Ni) + 5--50 & 5\% & 3--12 & ITO glass slides & NR \\
    Das \textit{et al.}\cite{das_boundaries_2015} & 2, 3.1, 4.8 $\mu$m & 10 & 10\% & 0--2 & Rectangular wells on Si wafers & Yes$^c$ \\
    Haroon \textit{et al.}\cite{haroon_coupled_2026} & 3 $\mu$m & 3--35 & 1--3\% & 0.5--3 & Glass slide, plasma cleaned & Yes$^d$ \\
    Liu \textit{et al.}\cite{liu_chemical_2024} & PS, SiO$_2$ (1--10 $\mu$m) & $\sim$20 & 5 wt\% & $\sim$5--20 & Coverslip, silicone oil--water interface & NR \\
    \bottomrule
  \end{tabularx}

  \vspace{1ex}
  \textsuperscript{\emph{Note:}} D = Diameter; NR = Not Reported; PS = Polystyrene. \\
  \textsuperscript{\emph{a}} Sonicated at 33 kHz for 5 min in 15\% w/v H$_2$O$_2$, then diluted. \\
  \textsuperscript{\emph{b}} Sonicated for 5 min in 15\% H$_2$O$_2$ and rested for 25 min to clean/activate. \\
  \textsuperscript{\emph{c}} Equal volumes of particles and 30\% H$_2$O$_2$ sonicated for 5 min, left for 20 min to activate. \\
  \textsuperscript{\emph{d}} Particles submerged in 15\% H$_2$O$_2$ for 30 min to activate.
\end{table*}

Despite the wide use of Janus particles as a model system, experimental protocols vary considerably between laboratories, and several parameters that affect speed are often uncontrolled or unreported. Table~\ref{tbl:literature_review} summarizes key parameters for a representative set of Pt/PS JP studies. Particle diameters range from 0.5 to 10~$\mu$m. Platinum thickness, where reported, ranges from 1 to 135~nm, a range over which particle behavior can change from active propulsion to orientational quenching\cite{haroon_coupled_2026, meng_effect_2022}. The substrate is usually described only as ``glass'', without the cleaning method, surface charge or wettability, although surface chemistry changes propulsion speed through slip-dependent phoretic coupling\cite{ketzetzi_slip_2020} and ions released from treated surfaces can suppress speed across the whole cell\cite{haroon_coupled_2026}. A pre-activation step, in which particles are exposed to H$_2$O$_2$ before imaging to condition the catalytic surface, is reported in only 4 of the 17 studies, each with a single fixed protocol and without testing how the activation conditions affect speed\cite{campbell_gravitaxis_2013, das_boundaries_2015, raj_direct_2025, haroon_coupled_2026}. Wang and Mallouk have called for more rigorous reporting of nano-swimmer trajectories and analysis\cite{wang_practical_2021}, but the upstream variables of catalyst conditioning and substrate preparation have received little systematic attention. Here we quantified how platinum activation time and H$_2$O$_2$ concentration set the propulsion speed, how batch age and storage change this response, and how plasma cleaning of the glass substrate suppresses propulsion. From these results we gained additional physical insight into the relevant propulsion mechanisms and also propose a minimum reporting standard for Pt/PS JP speed experiments.

\section{Experimental Methods}
Full procedures are given in the Supporting Information. The important details are summarized below.

\subsection{Janus particles} Sulfate latex PS particles (3~$\mu$m, Fisher Scientific, Lot \#2855924) were spin-coated as monolayers on silicon wafers (Figure~\ref{fig:si-monolayer-polystyrene}) and coated with 7~nm of platinum by electron-beam evaporation. Two batches made from the same PS lot were used: an aged batch (Pt deposited in July 2024) and a fresh batch (March 2026). All dose--response measurements used the aged batch. Substrate experiments used the fresh batch after pre-activation. The pre-activation condition for the substrate experiments was 30 wt./vol\% for 10~min.

\subsection{Pre-activation} 
\begin{figure}
    \centering
    \includegraphics[width=1\linewidth]{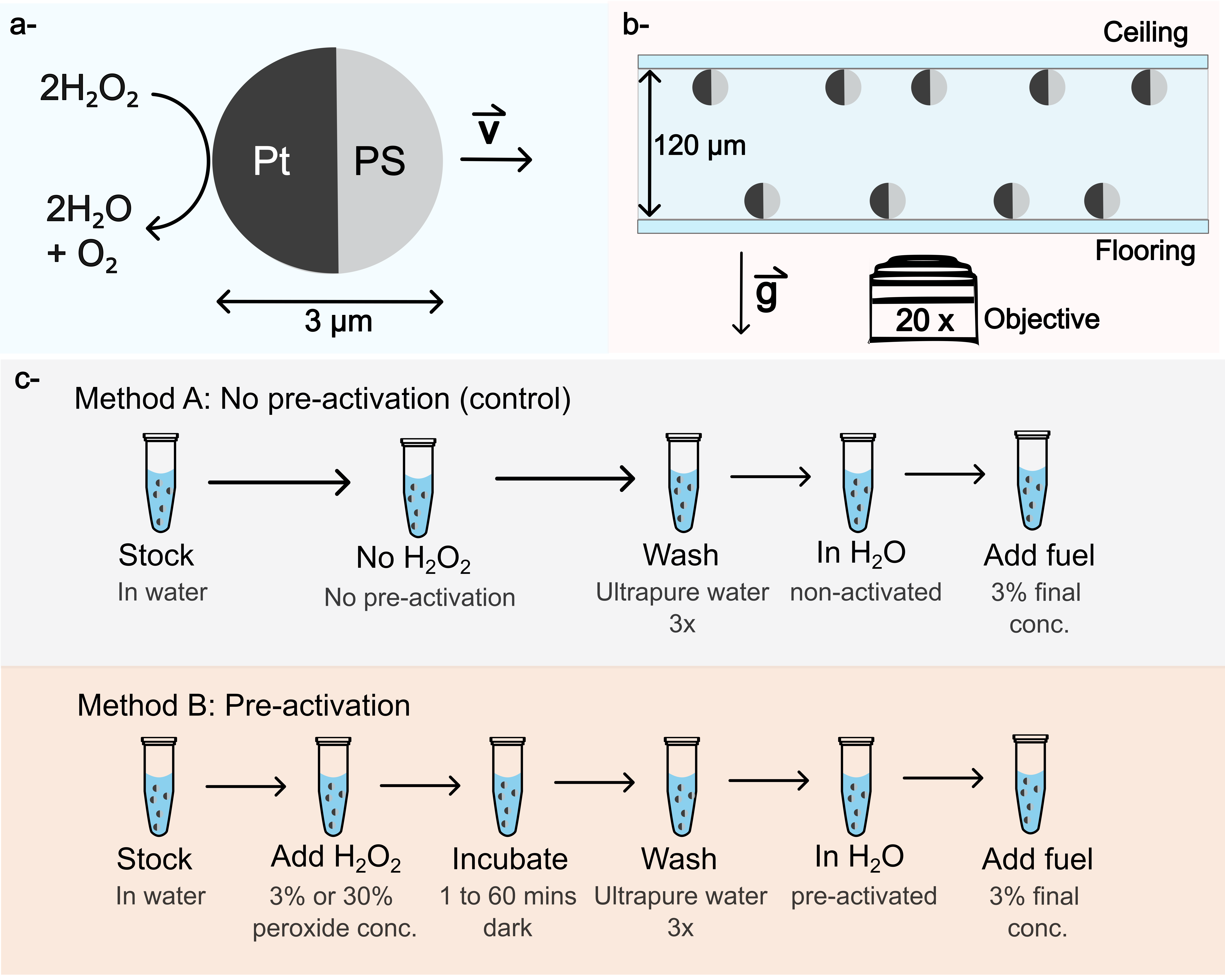}
    \caption{Janus particle system, imaging geometry and pre-activation protocols. (a) Catalytic decomposition of H$_2$O$_2$ on the platinum hemisphere propels a 3~$\mu$m Janus particle toward its polystyrene face. (b) Particles slide along the lower and upper boundaries of a 120~$\mu$m fluid cell, referred to as flooring and ceiling states. (c) Pre-activation protocols. In Method A the particles first contact peroxide when fuel is added to the assembled cell. In Method B the particles are incubated in 3 or 30 wt./vol\% H$_2$O$_2$ in the dark for 1 to 60~min, washed three times in water, and re-suspended before fuel is added.}
    \label{fig:intro-activation-method}
\end{figure}

Particle aliquots were incubated in 3 or 30 wt./vol\% H$_2$O$_2$ in the dark for 1 to 60~min. Particles were washed three times by centrifugation (Figure~\ref{fig:intro-activation-method}c). Non-activated controls followed the same procedure with water in place of H$_2$O$_2$.

\subsection{Substrate and fluid cell preparation} Precleaned glass slides were used as received for the pre-activation study. For the substrate study, slides were either used as received  (untreated, N.P.C.) or sonicated in isopropanol and acetone and then air-plasma cleaned for 2~min (P.C.). Cells were assembled with 120~$\mu$m or 2600~$\mu$m high adhesive spacers (9~mm diameter) and filled with particles in 3 wt./vol\% H$_2$O$_2$.

\subsection{Imaging and analysis} Each cell was imaged with a 20$\times$ objective at 10~fps, recording 100~s videos alternately at the ceiling and flooring planes (five each per cell, three cells per condition). Tracks were extracted with TrackMate, and the speed $v$ of each particle was obtained by fitting $\mathrm{MSD}(\tau) = 4D\tau + v^2\tau^2$ for $\tau \le 1$~s.  Conditions were compared with Welch's ANOVA and the Games--Howell post hoc test, which do not assume equal variances or sample sizes\cite{games1976pairwise}.

\subsection{Surface characterization} XPS survey spectra of untreated and plasma-cleaned glass were recorded on a PHI 5000 VersaProbe II. Ions released by the glass were measured as the conductivity of ultra-pure water in contact with five slides (0.40~mm$^{-1}$ glass area per volume) for 20~min.

\section{Results and Discussion}
\subsection{Catalyst pre-activation}\label{sec:preactivation}

\subsubsection{Pre-activation of the platinum cap increases propulsion speed}

\begin{figure}
    \centering
    \includegraphics[width=1\linewidth]{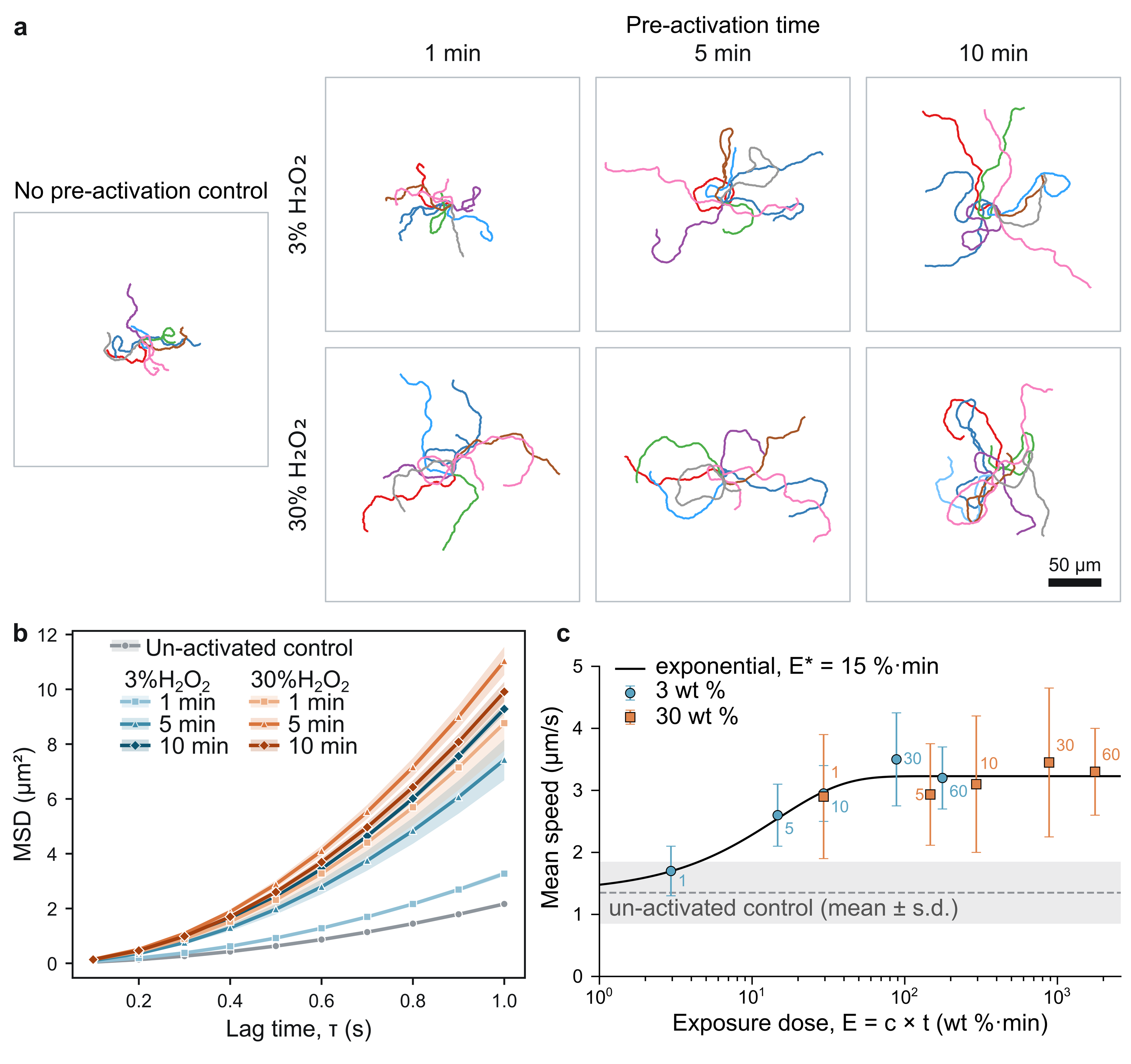}
    \caption{Pre-activation of the platinum cap increases propulsion speed. (a) Representative 100~s trajectories of non-activated particles and of particles pre-activated for 1, 5 and 10~min in 3 and 30 wt./vol\% H$_2$O$_2$; each color represents a distinct particle. (b) Ensemble mean-squared displacement against lag time for the same conditions, with shaded bands showing one standard deviation. (c) Mean speed against pre-activation dose, with incubation times in minutes beside each point. The solid line is the fit of Eq.~\ref{eq:activation_dose} with $E^* = 15$ wt./vol\%$\cdot$min. The shaded band is the non-activated control, mean $\pm$ one standard deviation. Error bars represent one standard deviation.}
    \label{fig:activation-main}
\end{figure}

Particles that were not pre-exposed to peroxide before the experiment traveled a shorter distance in 100~s than particles that were pre-exposed, i.e. pre-activated, as shown by the trajectories in Figure~\ref{fig:activation-main}a. After pre-exposure to 3 wt./vol\% H$_2$O$_2$, the distance traveled grew as the incubation time increased from 1 to 10~min, whereas after pre-exposure to 30 wt./vol\% it was similar for all incubation times. The mean-squared displacements showed a similar trend (Figure~\ref{fig:activation-main}b). Pairwise comparisons showed that in the 3 wt./vol\% series the mean speed rose significantly from 1 to 5 to 10~min, while the 1, 5 and 10~min conditions at 30 wt./vol\% did not differ from one another (Figure~\ref{fig:si-activation-time}). When mean speed was plotted against the pre-activation exposure dose, the product of peroxide concentration and incubation time, the data for both concentrations fell on one curve (Figure~\ref{fig:activation-main}c). Conditions with similar exposure doses gave similar speeds. For example, 3 wt./vol\% for 10~min and 30 wt./vol\% for 1~min are both exposure doses of 30 wt./vol\%$\cdot$min and gave similar mean speeds (Figure~\ref{fig:si-activation-time}). The data were described empirically by an exponential saturation:
\begin{equation}
    v = v_0 + (v_\infty - v_0)\left[1 - \exp\!\left(-\frac{E}{E^*}\right)\right],
    \label{eq:activation_dose}
\end{equation}
where $v_0$ is the speed of non-activated particles, $v_\infty$ the plateau speed, $E = c\,t$ the nominal exposure dose and $E^*$ the characteristic exposure dose. The fit gave $v_\infty \approx 3.2$~$\mu$m/s and $E^* = 15 \pm 3$~wt./vol\%$\cdot$min, corresponding to about 5~min at 3 wt./vol\% and 30~s at 30 wt./vol\%. Beyond about $2E^*$, mean speeds lay between 2.9 and 3.5~$\mu$m/s with no systematic dependence on exposure dose, up to the largest exposure dose tested. The relation is empirical, but it shows that cumulative exposure, or exposure dose, is a useful experimental variable.

A saturating response is consistent with a finite population of surface sites being converted to a conditioned state on a catalyst surface. The 2.4-fold difference between our non-activated control (1.36~$\mu$m/s) and the plateau (3.2~$\mu$m/s) covers most of the range reported for comparable Pt/PS systems in Table~\ref{tbl:literature_review}. Das et al. reported 1.5--2.1~$\mu$m/s for 2.4~$\mu$m PS with 9~nm Pt in 3 v/v\% H$_2$O$_2$\cite{das_floor-_2020}, and Issa et al. reported 0.2--1.4~$\mu$m/s for 3~$\mu$m PS with 20~nm Pt in 3\% H$_2$O$_2$\cite{issa_engineered_2023}. Neither study reported an activation step. They also differ from ours in particle size and cap thickness, so pre-activation is not the only difference, but on its own it can produce a spread of this size.

\subsubsection{Activation during measurement is not observed}
During a measurement the particles sat in 3 wt./vol\% H$_2$O$_2$ for 20--25~min, a nominal exposure dose of 60--75 wt./vol\%$\cdot$min, or 4--5$E^*$. If exposure alone activated the platinum, non-activated particles should have approached the plateau of $\sim$3~$\mu$m/s during the recording. A weighted linear fit of speed against time gave $+0.014 \pm 0.007$~$\mu$m~s$^{-1}$~min$^{-1}$ for the control (Figure~\ref{fig:si-with-time-control-activated}), a rise of only $\sim$0.35~$\mu$m/s over 25~min, compared with the $\sim$1.8~$\mu$m/s rise expected for full activation. Pre-activated particles slowed slightly over the same period (Figure~\ref{fig:si-with-time-control-activated}). Exposure in the cell therefore did not reproduce pre-activation. We suggest that activation requires two steps: conditioning of the platinum surface by peroxide, followed by removal of the peroxide solution. In the first step, peroxide exposure forms chemisorbed oxygen on the platinum. H$_2$O$_2$ decomposition on Pt proceeds through a Pt(O) intermediate, and a surface that already carries this oxygen reacts faster at first because the rate-limiting step has already taken place\cite{serra2018mechanism}. In the second step, the three wash cycles remove the peroxide solution together with any material that the exposure has released from the particles. Several preparation steps (spin coating, platinum deposition and sonication from the wafer) can leave organic or ionic residues on the cap, and adsorbed molecules are known to slow catalytic Janus particles strongly\cite{greydanus2022probing}. If peroxide exposure releases such residues, they can re-adsorb on the platinum unless the solution is replaced. In a sealed fluid cell the particles were exposed to peroxide but never washed, which would explain why non-activated particles did not reach the speed of pre-activated particles during the measurement. Consistent with removal of surface species, survey XPS of a platinum film showed that Na (7.8 at.\%) fell below detection after a brief exposure to 0.1\% H$_2$O$_2$ (Figure~\ref{fig:SI_Pt_XPS}). The O signal did not increase, so these data do not show the first step directly. Comparing washed and unwashed particles after peroxide exposure would test this idea, but residual peroxide in unwashed samples would change the fuel concentration and confound the comparison.

\subsubsection{The activation response depends on particle batch history}
\begin{figure}
    \centering
    \includegraphics[width=1\linewidth]{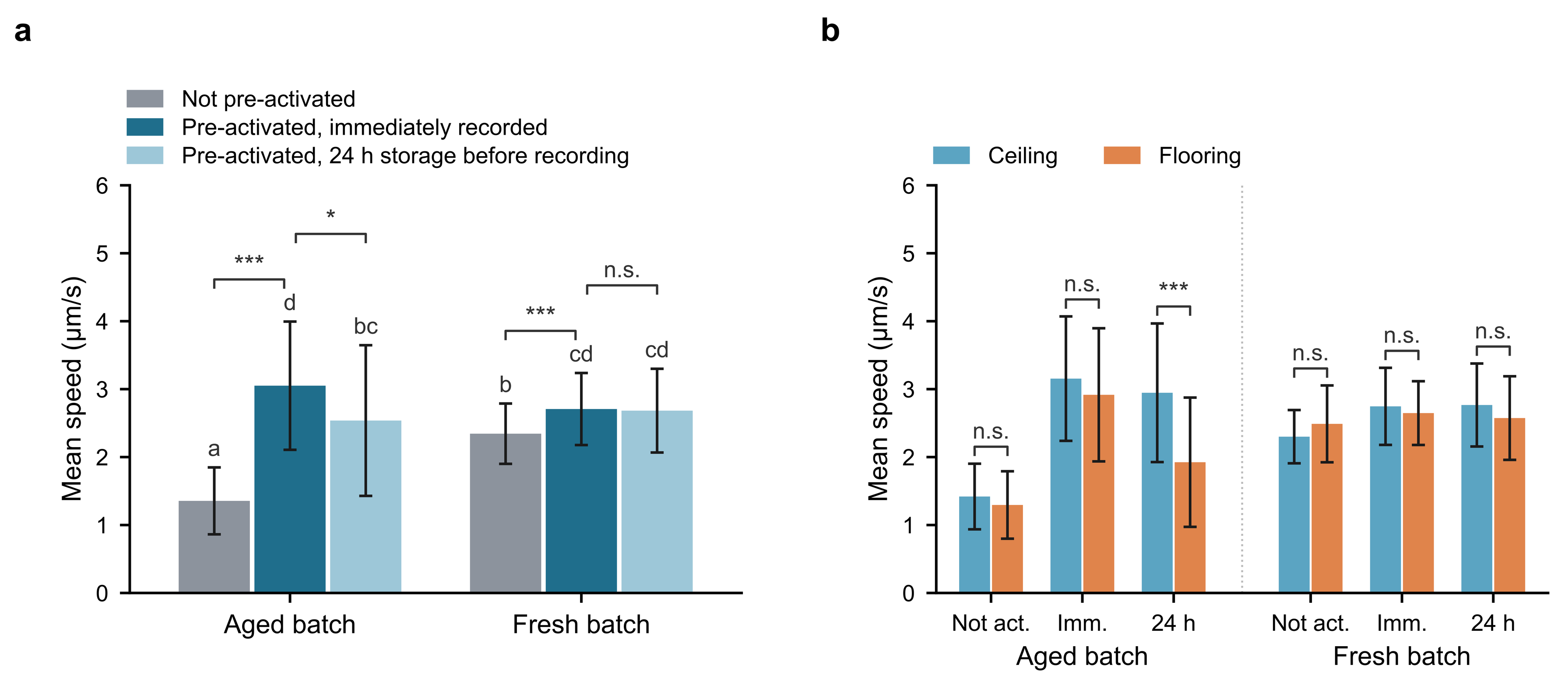}
    \caption{The activation response depends on batch history. (a) Mean speed of an aged and a fresh batch without pre-activation, immediately after pre-activation in 30 wt./vol\% H$_2$O$_2$ for 10~min, and after the same pre-activation followed by 24~h of storage in water; ceiling and flooring particles pooled ($n = 63$--$194$). Conditions sharing a letter do not differ (Games--Howell, $\alpha = 0.05$). (b) The same measurements resolved by imaging plane; brackets compare ceiling and flooring within each condition (Welch's $t$-test, Holm-adjusted). Bars are means, error bars are one standard deviation. *** $p < 0.001$; * $p < 0.05$; n.s., not significant.}
    \label{fig:activation-re-activation-aged-fresh}
\end{figure}

Figure~\ref{fig:activation-re-activation-aged-fresh}a compares two batches of nominally identical particles. An aged batch had Pt deposited in July 2024 and a fresh batch had Pt deposited in March 2026, both with 7~nm Pt on the same lot of PS particles. In the aged batch, pre-activation (30 wt./vol\%, 10~min) raised the mean speed from $1.36 \pm 0.49$ to $3.12 \pm 1.07$~$\mu$m/s, a factor of 2.3 ($p < 0.001$). In the fresh batch the same treatment raised it by only 16\%, from $2.34 \pm 0.44$ to $2.71 \pm 0.53$~$\mu$m/s ($p < 0.001$). Before pre-activation the fresh batch was measured to be 1.7 times faster than the aged batch, and after pre-activation the two batches could not be distinguished ($p = 0.065$). Pre-activation therefore appeared to restore activity that the aged batch had lost rather than create new activity. Interestingly, Liu et al. made a related observation: the speed enhancement of Pt/PS motors at an oil--water interface fell from 4.7-fold to 2.2-fold after 3~$\mu$m particles were stored for one month\cite{liu_chemical_2024}. Their measurements differ from ours, but both identify time since fabrication as a variable that changes measured speed.

After 24~h of storage in water following pre-activation, the aged batch moved at $2.54 \pm 1.11$~$\mu$m/s, 19\% below the value measured immediately after pre-activation ($p = 0.015$) and still 1.9 times faster than non-activated particles ($p < 0.001$). The fresh batch did not lose speed on storage ($2.68 \pm 0.62$ versus $2.71 \pm 0.53$~$\mu$m/s, $p = 1.00$), and after storage the two batches were again indistinguishable ($p = 0.87$). Most of the activation benefit therefore survived 24~h of storage. In case of stored aged batch the slower flooring particles are the part of the population that lost activity during storage and no longer propelled strongly enough to reach the ceiling. Slower particles populating the flooring rather than ceiling is shown in our previous study \cite{haroon_coupled_2026}. One possible cause for the slowdown of particles is re-adsorption onto the platinum of species released from the aged particles into the storage water, which would also fit the large activation response of the aged batch. Adsorbed contaminants have been shown to reduce the rate of peroxide decomposition on platinum, which could in turn reduce the speed of active particles\cite{greydanus2022probing}.

\begin{figure}
    \centering
    \includegraphics[width=1\linewidth]{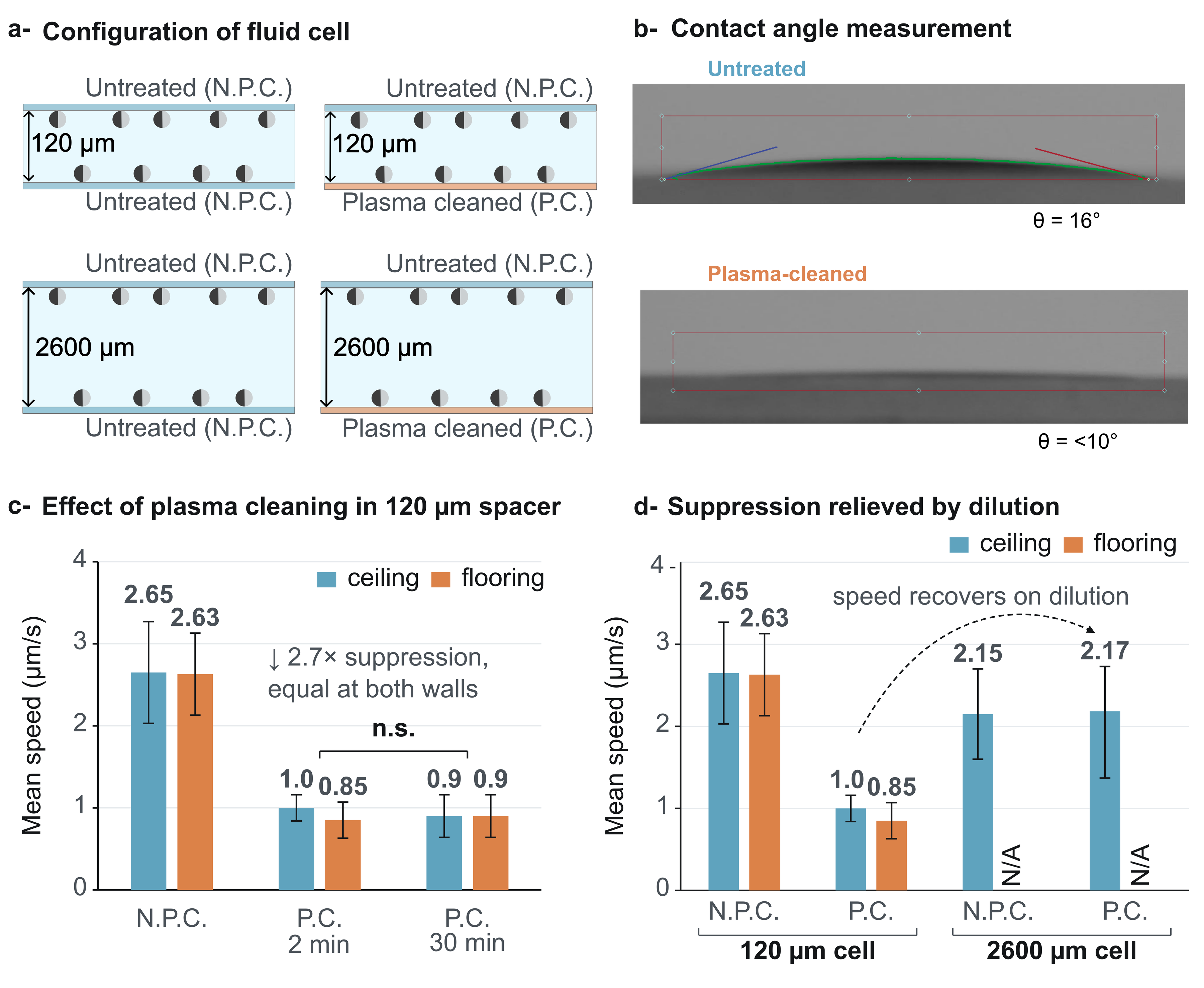}
    \caption{Plasma cleaning of the substrate suppresses propulsion at both boundaries. (a) Cell configurations with untreated (N.P.C.) or plasma-cleaned (P.C.) bottom slides at 120~$\mu$m and 2600~$\mu$m spacer heights. (b) Sessile-drop contact angles of untreated and plasma-cleaned glass. (c) Mean speed in the 120~$\mu$m cell on untreated glass and on glass plasma cleaned for 2~min and 30~min; blue, ceiling; orange, flooring. (d) Mean speed in 120~$\mu$m and 2600~$\mu$m cells. Flooring was not observed in the 2600~$\mu$m cells and is marked N/A. Bars are means, error bars one standard deviation.}
    \label{fig:plasma-cleaning-main}
\end{figure}

\subsection{Substrate plasma cleaning}\label{sec:plasma}
\subsubsection{Plasma cleaning suppresses propulsion}

Replacing the untreated bottom slide with a plasma-cleaned slide lowered the mean speed from $2.65 \pm 0.62$ to $1.00 \pm 0.16$~$\mu$m/s at the ceiling and from $2.63 \pm 0.50$ to $0.85 \pm 0.22$~$\mu$m/s at the flooring, about a 2.7-fold reduction (Figure~\ref{fig:plasma-cleaning-main}c). Plasma cleaning for 30~min instead of 2~min gave no further reduction. The suppression was the same at both walls even though only the bottom slide was treated. Wettability is known to affect propulsion near a wall: Ketzetzi et al. found that the speed scales as $V \propto (1 + \cos\theta)^{-3/2}$ with the substrate contact angle\cite{ketzetzi_slip_2020}. Plasma cleaning reduced the water contact angle from $16^{\circ}$ to below $10^{\circ}$ (Figure~\ref{fig:plasma-cleaning-main}b), for which that relation predicts a speed reduction of at most 3\%, far smaller than the 62\% observed experimentally. Hydrodynamic slip associated with wettability was therefore not the main cause of the suppression. In a taller, 2600~$\mu$m cell, treating the bottom slide did not reduce the speed of ceiling particles (Figure~\ref{fig:plasma-cleaning-main}d). Increasing the fluid volume while keeping the treated area the same therefore relieved the suppression. Because treating the bottom slide slowed particles at the opposite, untreated wall of the 120~$\mu$m cell, a mechanism acting through the solution is more plausible than a boundary effect.

\begin{figure}
    \centering
    \includegraphics[width=1\linewidth]{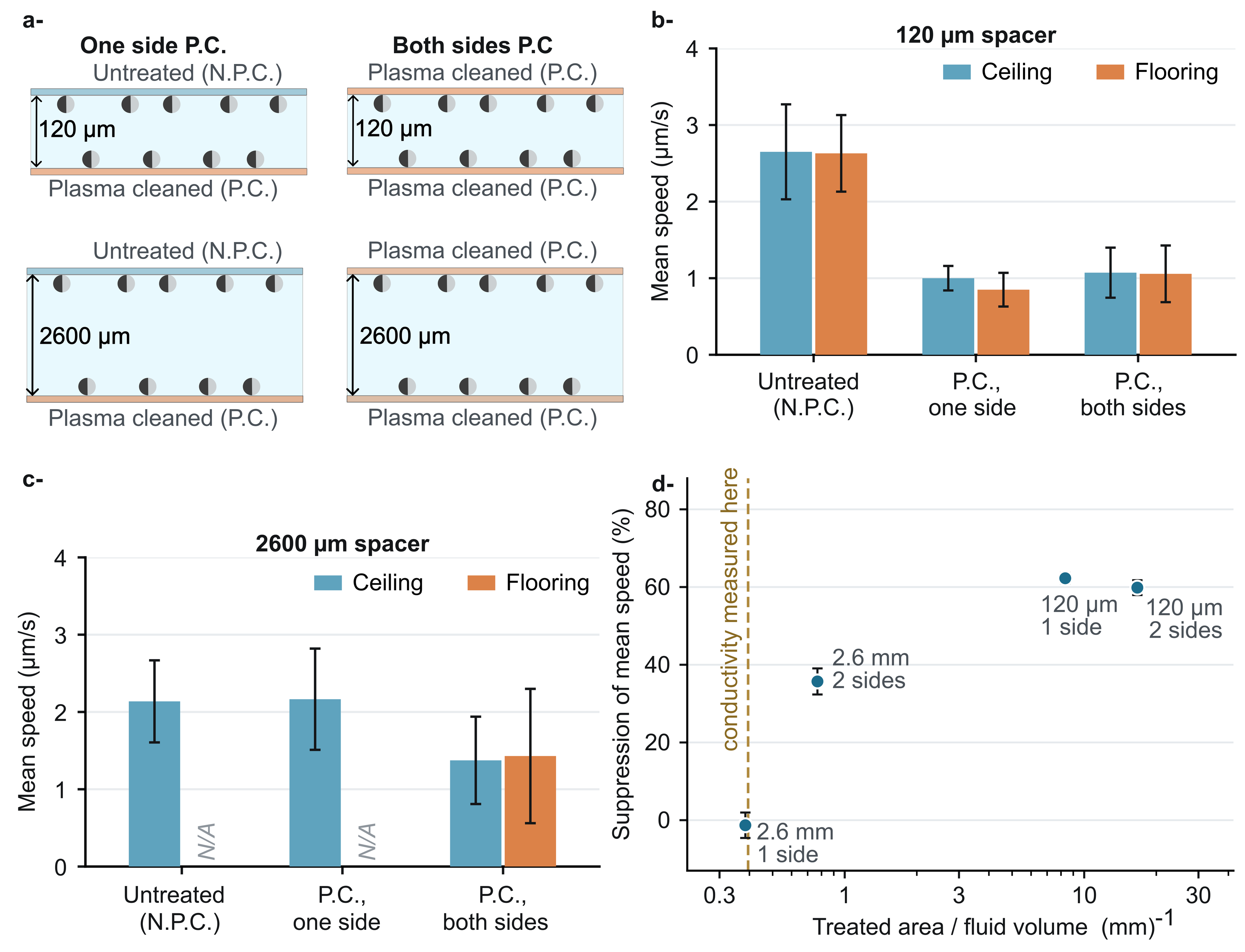}
    \caption{Suppression increases with treated glass area per unit fluid volume. (a) Cell configurations with one or both walls plasma cleaned, at 120~$\mu$m and 2600~$\mu$m spacer heights. (b) Mean speed by imaging plane in the 120~$\mu$m cell for untreated, one-wall and two-wall treated configurations. (c) The same comparison in the 2600~$\mu$m cell. Flooring was not observed for the untreated and one-wall 2600~$\mu$m cells and is marked N/A. (d) Suppression of the mean speed against treated glass area per unit fluid volume, $N/h$, for $N$ treated walls separated by a gap $h$. The vertical marker shows the area-to-volume ratio of the conductivity measurement in Figure~\ref{fig:xps-contact-angle-mechanism}b.}
    \label{fig:area-volume-scaling}
\end{figure}

\subsubsection{Suppression increases with treated area per unit volume}
Treating both walls of the 2600~$\mu$m cell instead of one suppressed the speed by 36\% (Figure~\ref{fig:area-volume-scaling}c). In the 120~$\mu$m cell, however, treating both walls gave no further suppression than treating one. If the treated glass acts as a source of dissolved species, the relevant variable is the amount released per unit volume of fluid, and the suppression should weaken when the same treated area faces a larger volume. For a parallel-plate cell with $N$ treated walls separated by a gap $h$, the treated area per unit volume is $N/h$, independent of the well diameter. $N/h$ is therefore the treated surface-area-to-volume ratio of the cell. Across the four configurations in Figure~\ref{fig:area-volume-scaling}, $N/h$ spans 0.385 to 16.7~mm$^{-1}$, and the suppression rose from $-1.3$\% at 0.385~mm$^{-1}$ to 35.7\% at 0.769~mm$^{-1}$ and 62.3\% at 8.33~mm$^{-1}$, then leveled off at 59.5\% at 16.7~mm$^{-1}$. Suppression was therefore absent at low treated area per volume and appeared to saturate at high values. Although only four configurations were tested, this dependence indicates that the suppression is caused by changes in the solution that scale with the area of treated glass in contact with it.

\begin{figure}
    \centering
    \includegraphics[width=1\linewidth]{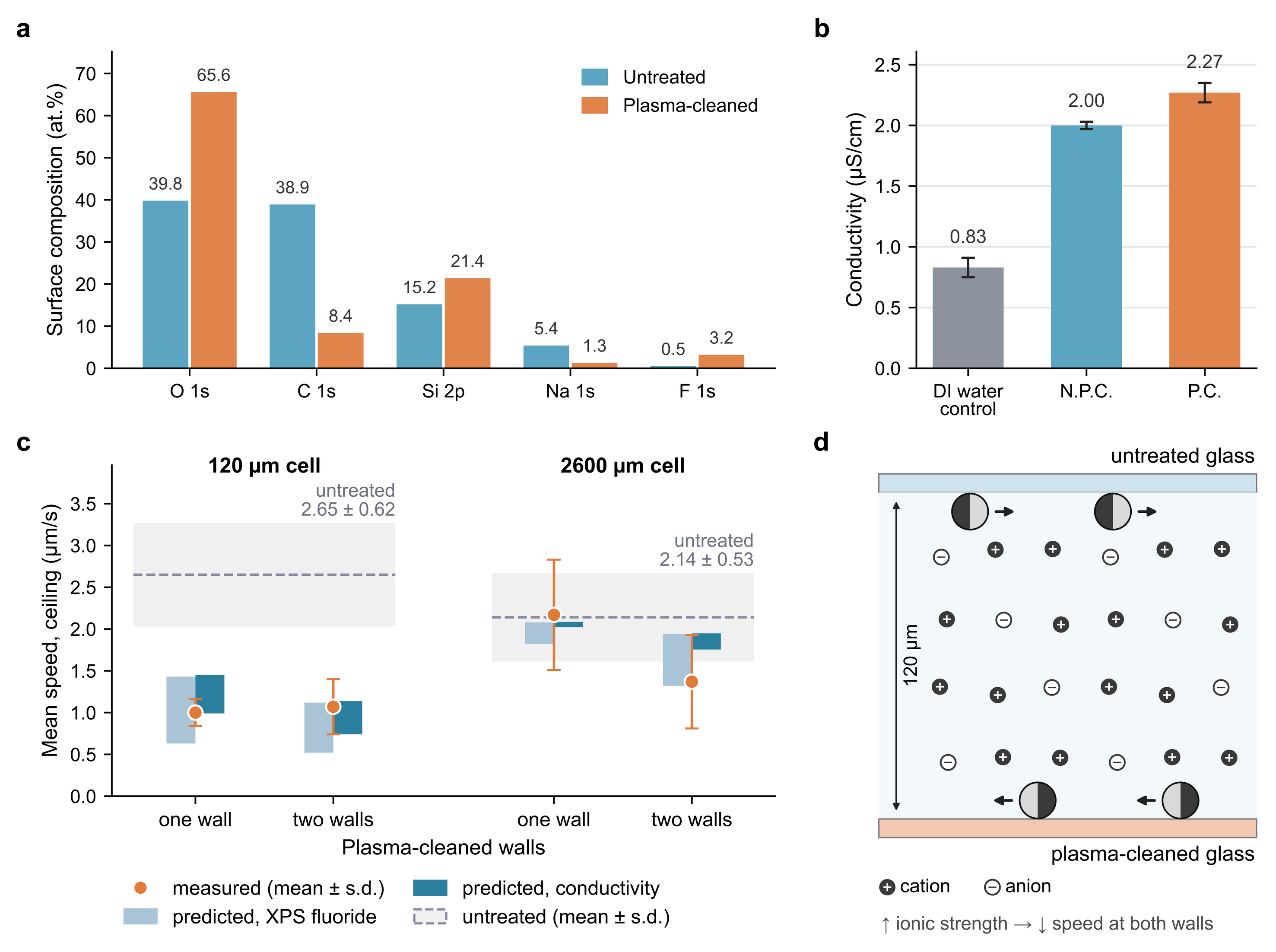}
   \caption{Surface chemistry of plasma-cleaned glass and solution-mediated suppression. (a) Surface composition of untreated and plasma-cleaned glass from XPS survey spectra (Figure~\ref{fig:SI_glass_XPS}). (b) Conductivity of ultra-pure water after 20~min in contact with untreated (N.P.C.) or plasma-cleaned (P.C.) glass, with a water control; mean $\pm$ s.d.\ of triplicates. (c) Measured mean speed of ceiling particles (orange; mean $\pm$ s.d.) and speeds predicted from Eq.~\ref{eq:vpred} with $\Delta\sigma$ from the conductivity increment in (b) scaled to the treated area per volume of each cell (dark bars) or from complete dissolution of the fluoride detected by XPS (light bars). Predicted bars span $\sigma_0 = 2$--$4~\mu$S/cm and $v_\mathrm{res} = 0.4$--$0.6~\mu$m/s and, for the XPS estimate, both limits of the fluoride inventory. Dashed lines and grey bands represent untreated cell of the same geometry.
  (d) Proposed mechanism: ions released from plasma-cleaned glass raise the ionic strength throughout the cell and lower the speed at both walls.}
    \label{fig:xps-contact-angle-mechanism}
\end{figure}

\subsubsection{Tentative identity of the released species}
XPS showed that plasma cleaning removed adventitious carbon from the glass (38.9 to 8.4 at.\%; Figure~\ref{fig:xps-contact-angle-mechanism}a). Plasma cleaning also lowered Na from 5.4 to 1.3 at.\% and raised F from 0.5 to 3.2 at.\%. Removing carbon alone would raise the non-carbon fraction from 61 to 92 at.\%, which would scale every other element by about 1.5 and bring F to only 0.75 at.\% and Na to 8.1 at.\%. The measured values therefore reflect a real gain of fluorine (3.2 at.\%) and loss of sodium (1.3 at.\%), not just renormalization due to carbon loss. A likely source of the fluorine is fluoroelastomer or fluoropolymer components of the plasma chamber seal. At the same time, the conductivity of water in contact with plasma-cleaned slides reached $2.27 \pm 0.08$~$\mu$S/cm, compared with $2.00 \pm 0.03$~$\mu$S/cm for untreated slides and $0.83 \pm 0.08$~$\mu$S/cm for the water control (Figure~\ref{fig:xps-contact-angle-mechanism}b). Although 0.27 ~$\mu$S/cm difference seem small between untreated and plasma cleaned slides, it is measured at the lower surface-area to volume ratio of 0.4 ~mm$^{-1}$. As the surface-area to volume ratio increases up to 40-fold in smaller fluid cell, the difference in conductivity of the solution increases with it. The control was higher than freshly dispensed 18.2~M$\Omega\cdot$cm water (0.055~$\mu$S/cm) because it equilibrates with atmospheric CO$_2$. Both glasses released ions, but plasma-cleaned slides released more ions into water than non-plasma-cleaned slides. The ions released by the two glasses, however, are not expected to affect the particle speed in the same way. In water, soda-lime glass exchanges network Na$^+$ for H$_3$O$^+$, so its leachate is alkaline, essentially NaOH\cite{bunker1994molecular}. Brown and Poon found that adding up to 0.1~mM NaOH does not significantly slow Pt/PS swimmers, whereas pH-neutral salts do\cite{brown_ionic_2014}. XPS detected Na but no Cl on either glass (Figure~\ref{fig:SI_glass_XPS}). We therefore treat the leachate of untreated glass as NaOH-like, with little effect on speed.

On the other hand, the additional conductivity from plasma-cleaned glass is more likely due to fluoride than to sodium, because Na decreased and F increased on the plasma-cleaned surface. The F~1s binding energy (685.1~eV) and modified Auger parameter (1339.5~eV) are those of an ionic metal fluoride, not of a fluorocarbon (F~1s 688--689~eV; Figure~\ref{fig:SI_glass_XPS}), and the C~1s spectrum showed no C--F component\cite{moulder1992handbook,naumkin_nist_xps}. Fluorine exceeded Na 2.5-fold, and the Na Auger parameter did not change, so the fluoride is not NaF. Instead, the Ca~2p$_{3/2}$ peak shifted from 347.1 to 347.8~eV, the value reported for CaF$_2$\cite{moulder1992handbook}, and Na + 2Ca balanced F within the accuracy of survey quantification. Fluorine was not part of the process gas (air), and the plasma cleaner's chamber and sample shelf are quartz, so the fluorine most likely came from the fluorinated vacuum grease on the door seal. The manufacturer recommends a perfluoropolyether (PFPE) grease for this O-ring. In the plasma, PFPE decomposes and its fluorine reacts with exposed surfaces to form metal fluorides\cite{chaug1993reactions}.
We therefore propose that plasma cleaning introduces a calcium-associated, CaF$_2$-like fluoride salt. Because the Ca shift was measured on survey spectra of an insulator referenced to adventitious carbon, whose amount changes on plasma cleaning, we regard the calcium assignment as tentative. On contact with the fuel, such a layer could release Ca$^{2+}$ and F$^-$, a pH-neutral salt.

\subsubsection{A conductivity model reproduces the measured speeds}\label{sec:model}

A released pH-neutral salt is expected to slow the particles. For self-electrophoretic propulsion the reaction drives an ionic current through the surrounding fluid, the resulting field scales inversely with solution conductivity\cite{paxton2006catalytically,moran2014role}. Brown and Poon tested this scaling for Pt/PS Janus particles by adding NaCl or KBr to 10\% H$_2$O$_2$\cite{brown_ionic_2014}. The speed was linear in the inverse conductivity but did not fall to zero: at high conductivity the particles kept a residual speed of about 1.3~$\mu$m/s.  The authors suggested that residual speed may come from neutral self-diffusiophoresis, a contribution that does not depend on conductivity. Their measurements also show how sensitive these particles are to salt. A 0.1~mM NaCl raises the conductivity of the fuel about fourfold from 3.9~$\mu$S/cm to 12 ~$\mu$S/cm\cite{brown_ionic_2014,melio2025motion}.

We estimated the speed in each plasma-cleaned cell with the same relation, written relative to the untreated cell of the same geometry:
\begin{equation}
    v = v_\mathrm{res} + \left(v_\mathrm{u} - v_\mathrm{res}\right)\frac{\sigma_0}{\sigma_0 + \Delta\sigma},
    \label{eq:vpred}
\end{equation}
where $v_\mathrm{u}$ is the measured speed in the untreated cell, $v_\mathrm{res}$ is the residual speed, $\sigma_0$ is the conductivity of the fuel, and $\Delta\sigma$ is the conductivity added by the plasma-cleaned glass, obtained by scaling the measured increment (0.27~$\mu$S/cm at 0.40~mm$^{-1}$) to the treated area per volume of the cell. We used a range for $\sigma_0$ between 2 and 4~$\mu$S/cm (Supporting Information). We estimated $v_\mathrm{res}$ for our particles in 3\% fuel from the residual measured by Brown and Poon in 10\% fuel. Because diffusiophoretic speed is proportional to the surface reaction rate\cite{howse_self-motile_2007}, we scaled their residual by the reaction rate at 3\% relative to 10\% H$_2$O$_2$, using the Michaelis--Menten kinetics that Howse et al.\ measured for Pt/PS particles\cite{howse_self-motile_2007}. This gives $0.45 \times 1.3 = 0.59~\mu$m/s. Our particles are also larger than those of Brown and Poon (3 versus 2~$\mu$m), and the speed of these particles decreases roughly in inverse proportion to their radius in this size range\cite{ebbens_size_2012}, which lowers the estimate to about 0.4~$\mu$m/s. We therefore used $v_\mathrm{res} = 0.4$--$0.6~\mu$m/s. Expressed as NaCl, $\Delta\sigma$ corresponds to about 45 and 90~$\mu$M in the 120~$\mu$m one- and two-wall cells, more than the conductivity of the fuel itself, but only about 1 and 5~$\mu$M in the 2600~$\mu$m cells. Independently, complete dissolution of the fluoride detected by XPS (6--18~$\mu$mol~m$^{-2}$) would add 0.28--0.84~$\mu$S/cm under the conditions of the conductivity measurement, consistent with the measured increment (calculations in supporting information). With either estimate of $\Delta\sigma$, Eq.~\ref{eq:vpred} reproduced the measured speeds within one standard deviation in all four cell geometries (Figure~\ref{fig:xps-contact-angle-mechanism}c and Table~\ref{tab:SI_velocity_prediction}). The residual term mattered mainly in the 120~$\mu$m cells, where $\Delta\sigma$ exceeded $\sigma_0$ several-fold and the conductivity-dependent part of the speed was largely suppressed. Without considering residual speed, the predicted speed in the two-wall cell fell to 0.40--0.70~$\mu$m/s, below the measured $1.07 \pm 0.33~\mu$m/s. With residual speed included, the second treated wall lowered the predicted speed only from 0.99--1.45 to 0.74--1.14~$\mu$m/s, a difference smaller than the scatter of the measurements ($1.00 \pm 0.16$ and $1.07 \pm 0.33~\mu$m/s) and consistent with the similar speeds measured in the two cells. The agreement does not depend on the exact residual speed. All four predictions remained within one standard deviation of the measurements for any $v_\mathrm{res}$ between 0.07 and 0.63~$\mu$m/s (Figure~\ref{fig:SI_residual}).

\subsection{Implications for reporting and propulsion mechanism}

The two preparation steps acted on different parts of the propulsion mechanism. Pre-activation changed the platinum surface and it raised the speed of the aged batch to that of the fresh batch. Whereas, exposure to peroxide in a sealed cell, without washing, did not increase velocity, suggesting that removal of adsorbates from solution is one requirement to increase speed of particles. These results suggest that the speed of catalytic active Janus articles was limited by the rate of peroxide decomposition on the platinum, which depends on the chemical history of the catalyst surface, including adsorbed residues and surface oxygen. Any contribution to propulsion that scales with the reaction rate is affected in the same proportion, so catalyst history acts as a multiplicative factor on the speed regardless of the propulsion mechanism.

Plasma cleaning, in contrast, left the particles unchanged and acted through the solution. The suppression of particles' speed increased with the amount of released ions and followed the inverse-conductivity relation measured by Brown and Poon\cite{brown_ionic_2014}. The suppression scaling with conductivity indicates that most of the speed in our untreated cells came from an ionic, conductivity-sensitive contribution, as expected for self-electrophoresis\cite{paxton2006catalytically,moran2014role}. The near-equal speeds in the one- and two-wall 120~$\mu$m cells, which Eq.~\ref{eq:vpred} reproduced only when a conductivity-independent term (residual velocity) was included, point to a second contribution that persists at high ionic strength, consistent with neutral self-diffusiophoresis\cite{brown_ionic_2014,howse_self-motile_2007}. Our data therefore support a mixed mechanism in which an ionic contribution dominates at low conductivity and a smaller neutral contribution remains when the ionic contribution is suppressed. Also, the type of salt affecting the conductivity is important too. Salt sensitivity alone cannot distinguish self-electrophoresis from ionic self-diffusiophoresis, since both weaken with added electrolyte, so we refer to the dominant contribution simply as ionic.

The contribution of multiple mechanisms of propulsion explains why preparation steps can matter as much as the nominal specification of an experiment. A measured speed reflects two separate factors: how fast the platinum turns over fuel, which depends on catalyst history, and how much of the resulting ionic driving force survives in the solution, which depends on its conductivity. If careful consideration to experimental details is not applied, both catalyst history and conductivity can be altered without noticing. In our 120~$\mu$m cells, ions released from less than 1~cm$^2$ of plasma-cleaned glass lowered the speed by about 60\%. Ions from other sources, such as dissolved CO$_2$, leachate from untreated glass, buffers or stabilizers in commercial peroxide, enter Eq.~\ref{eq:vpred} in the same way. Two studies with identical particles and fuel concentration can therefore report different speeds, and even reach different conclusions about the dominant mechanism, if their solutions differ in conductivity: the ionic contribution dominates in a low-conductivity cell, whereas the neutral contribution makes up a larger share of the speed in a high-conductivity one. Reporting the catalyst history together with the conductivity of the fuel as used in the cell would make speeds comparable across studies and allow the two contributions to be separated. Wang and Mallouk have argued for more rigorous reporting of nano-swimmer trajectories and analysis\cite{wang_practical_2021}. Our results extend that argument to the steps before the measurement. Table~\ref{tbl:reporting} lists the minimum set of parameters we recommend reporting for Pt/PS Janus particle experiments.
\begin{table*}[htbp]
  \footnotesize
  \caption{Recommended minimum reporting standard for Pt/PS Janus particle speed measurements}
  \label{tbl:reporting}
  \begin{tabularx}{\textwidth}{>{\raggedright\arraybackslash}p{0.22\textwidth} X}
    \toprule
    \textbf{Category} & \textbf{Report} \\
    \midrule
    Particle & Core material, diameter and supplier lot; nominal Pt thickness and the position at which it is defined \\
    Deposition & Method (sputter / e-beam), chamber atmosphere and base pressure, deposition rate, adhesion layer if used \\
    Batch history & Date of deposition, elapsed time to measurement, storage medium and temperature \\
    Pre-activation & Peroxide concentration, incubation time, mixing and light conditions, number of wash cycles and wash medium, and time between pre-activation and measurement; state explicitly if no pre-activation was used \\
    Fuel & Peroxide concentration, basis (wt./vol\%, v/v, M)\\
    Water & Type and conductivity of the water used for washing, storage and fuel \\
    Substrate & Material, cleaning sequence, plasma conditions if used, and measured contact angle \\
    Cell geometry & Spacer height, well diameter, fill volume, which walls were treated \\
    Timing & Elapsed time between cell assembly and first frame, and total recording duration \\
    Imaging & Objective, frame rate, pixel calibration, and which boundary was imaged (ceiling, flooring, or both) \\
    Analysis & MSD fitting window relative to $\tau_R$, track-length and displacement filters, statistical tests, and the number of independent preparations as distinct from the number of tracks \\
    \bottomrule
  \end{tabularx}
\end{table*}

\section{Conclusions}
This manuscript summarizes experiments and analysis in which we examined how two routine but rarely reported preparation steps, pre-activation of the platinum cap in H$_2$O$_2$ and plasma cleaning of the glass substrate, affect the speed of Pt/PS Janus particles. Neither step is part of the nominal specification of a Janus particle experiment, yet altering those steps changed the measured propulsion speed by factors of 2.4 and 2.7. Pre-activation followed a saturating dose response and appeared to restore activity that an aged batch lacked. Plasma cleaning of the glass substrate suppressed propulsion along both boundaries of the cell, even when only one boundary is plasma cleaned. The suppression of speed grew with treated area per unit fluid volume, and was absent in case of lowest surface-area to volume ratio. XPS and conductivity measurements suggested that plasma cleaning deposits an ionic fluoride that dissolves as a pH-neutral salt, and a conductivity-based model of the released ions reproduced the measured speeds in all four cell geometries. Together, these results indicate that the speed of these particles is set both by the catalytic activity of the platinum surface and by the conductivity of the surrounding solution, with an ionic, conductivity-sensitive contribution dominating and a smaller conductivity-independent contribution remaining at high ionic strength. Absolute speeds can therefore only be compared across studies if catalyst history and solution and container chemistry are reported along with particle size, cap thickness and fuel concentration. The reporting standard in Table~\ref{tbl:reporting} is offered toward that end.
\begin{suppinfo}
Detailed experimental methods; estimate of the speed in plasma-cleaned cells (Tables~\ref{tab:SI_scaling} and \ref{tab:SI_velocity_prediction}); optical micrographs of polystyrene monolayers (Figure~\ref{fig:si-monolayer-polystyrene}); mean speed versus pre-activation time with pairwise comparisons (Figure~\ref{fig:si-activation-time}); speed versus elapsed time in the fluid cell (Figure~\ref{fig:si-with-time-control-activated}); pairwise comparisons for aged and fresh batches by imaging plane (Figure~\ref{fig:aged-fresh-heatmap-p-values}); particle numbers and speeds for aged and fresh batches (Table~\ref{tab:SI_aged_fresh_desc}); XPS survey spectra of untreated and plasma-cleaned glass (Figure~\ref{fig:SI_glass_XPS}); XPS of the platinum film before and after peroxide exposure (Figure~\ref{fig:SI_Pt_XPS}); dependence of the predicted speeds on the residual speed (Figure~\ref{fig:SI_residual}). In this preprint, the Supporting Information follows the reference list.

\end{suppinfo}

\begin{acknowledgement}
The authors thank Dr. Zeev Porat for helpful discussions. This work was supported by the National Science Foundation (NSF) Award No. 2314405. AI tools were used to generate Python code for plotting the data and to proofread the manuscript.
\end{acknowledgement}

\bibliography{achemso-demo}

\clearpage
\onecolumn
\setcounter{section}{0}
\setcounter{figure}{0}
\setcounter{table}{0}
\setcounter{equation}{0}
\setcounter{page}{1}
\renewcommand{\thesection}{S\arabic{section}}
\renewcommand{\thefigure}{S\arabic{figure}}
\renewcommand{\thetable}{S\arabic{table}}
\renewcommand{\theequation}{S\arabic{equation}}
\renewcommand{\thepage}{S\arabic{page}}

\begin{center}
{\Large\bfseries Supporting Information}\\[1.5ex]
{\large Impact of Surface Preparation on the Propulsion of Catalytic Janus Colloids and Implications for Reproducibility}\\[1.5ex]
Muhammad Haroon, Mitchell Lau, and Christopher L. Wirth\\[0.5ex]
{\small Department of Chemical and Biomolecular Engineering, Case Western Reserve University, Cleveland, OH 44106}
\end{center}

\noindent Figure, table, equation and section numbers prefixed with ``S'' refer to this Supporting Information; unprefixed numbers refer to the main text. Reference numbers refer to the reference list of the main text.

\section{Detailed methods}

\subsection{Janus particle fabrication}
\begin{figure}
    \centering
    \includegraphics[width=1\linewidth]{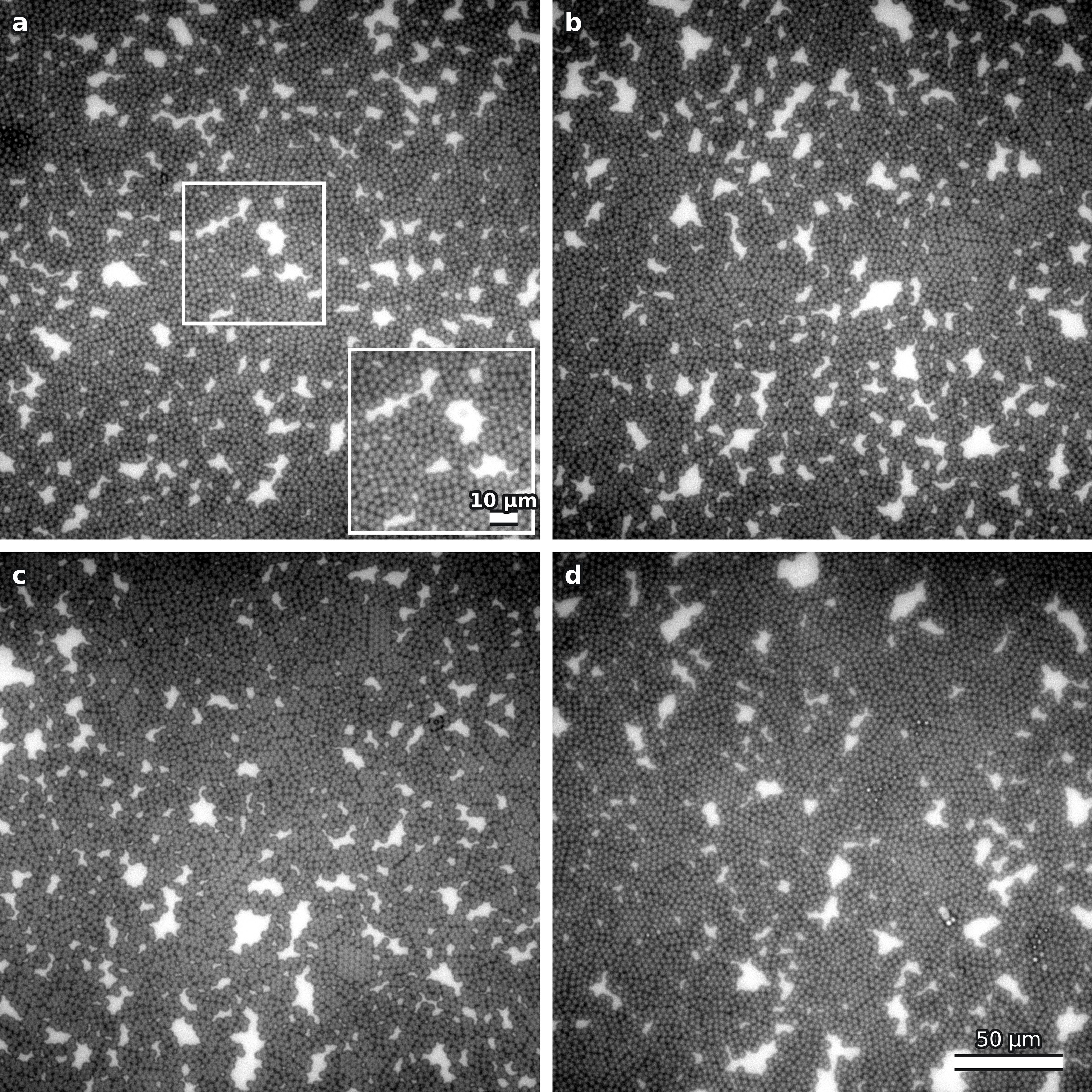}
    \caption{Optical micrographs of 3~$\mu$m sulfate latex polystyrene monolayers spin coated on 20~$\times$~20~mm silicon wafers before platinum deposition. (a--d) Four representative fields at 40$\times$. Particles form hexagonally close-packed domains separated by grain boundaries; bright regions are uncoated wafer. Coverage exceeds 80\%. The boxed region in (a) is enlarged in the inset. Scale bars: 50~$\mu$m (main panels), 10~$\mu$m (inset).}
    \label{fig:si-monolayer-polystyrene}
\end{figure}

Sulfate latex polystyrene particles (3~$\mu$m diameter, 8\% wt./vol, Fisher Scientific, Lot \#2855924) were sonicated for 30~min, and 1.5~mL was centrifuged (Thermo Scientific Sorvall Legend Micro) at 3000~$\times$~g for 10~min. The supernatant was replaced with ultra-pure water (18.2~M$\Omega\cdot$cm, Arium Mini, Sartorius), and the wash was repeated five times. A 9 wt./vol\% suspension gave the best monolayers ($>$80\% monolayer coverage). A 20~mm $\times$ 20~mm silicon wafer (Ted Pella) was plasma cleaned for 2~min, and 20~$\mu$L of suspension was spin coated (Ossila) at 500~rpm for 5~s and then 6000~rpm for 30~s. Monolayers were inspected at 40$\times$ (Olympus BX51WI; Figure~\ref{fig:si-monolayer-polystyrene}), and the wafers were fixed to glass slides with thermal tape for platinum deposition.

Platinum (7~nm nominal thickness at the pole) was deposited by electron-beam evaporation (Angstrom Engineering EvoVac) at 2~\AA/s and a base pressure of about $10^{-7}$~Torr, without an adhesion layer. Thickness was calibrated against a 30~nm reference film on glass measured by stylus profilometry. No significant flaking of the platinum and no uncoated particles were observed.

Particles were released from the wafers by sonication in ultra-pure water for 0.5--1~h, then washed five times by centrifugation at 3000~$\times$~g for 10~min and resuspension. A working suspension (20~$\mu$L of concentrate in 1000~$\mu$L water) gave about 10 particles per field of view (341~$\mu$m $\times$ 408~$\mu$m). The aged batch (Pt deposited July 2024) and the fresh batch (March 2026) were stored as aqueous suspensions in 1.5~mL vials in a refrigerator at 2--8~$^\circ$C.

\subsection{Pre-activation}
The stock suspension was vortexed and 20~$\mu$L aliquots were transferred to 1.5~mL vials. Pre-activation in 3\% or 30\% peroxide was performed as described below.

\textit{3\% activation.} 100~$\mu$L of 30\% H$_2$O$_2$ (Thermo Fisher Scientific, Certified ACS, Lot 242682) and 900~$\mu$L of DI water were added (2.94 wt./vol\% after dilution). The vial was vortexed and kept in the dark (inside the closed centrifuge) for the activation time, then vented, vortexed and centrifuged for 10~min at 2500~$\times$~g. About 90\% of the liquid was removed with a 1000~$\mu$L and then a 100~$\mu$L pipette and replaced with DI water, and the vial was vortexed; this wash was done three times in total. The exposure dose in Figure~\ref{fig:activation-main} of the main text counts only the incubation time and excludes the 10~min centrifugation, during which the particles sediment into a pellet and are no longer dispersed in the peroxide.

\textit{30\% activation.} 1000~$\mu$L of 30\% H$_2$O$_2$ was added (29.4 wt./vol\% after dilution). After incubation, the vented and vortexed suspension was split into four 250~$\mu$L portions, each diluted with 750~$\mu$L of DI water and washed three times as above.

After washing, the particles were suspended in DI water. Three tenfold dilutions left $\approx$0.003 (3\% protocol) and $\approx$0.007 wt./vol\% (30\% protocol) residual peroxide. Concentrated H$_2$O$_2$ is a strong oxidizer and its decomposition on platinum releases O$_2$, so closed vials were vented before mixing and centrifugation.

\subsection{Substrate preparation}
For the pre-activation study, precleaned glass slides (Fisherbrand Premium Plain) were used after removing dust with compressed air. For the substrate study, slides were sonicated in isopropanol (15~min), rinsed with ultra-pure water, dried with compressed air, and sonicated in acetone (15~min). Slides were rinsed again with water and dried with compressed air. For plasma cleaning condition slides were then plasma cleaned (Tergeo, air plasma, 2~min). Slides that were air-dusted only are called untreated or non-plasma-cleaned (N.P.C.), and slides that also received the plasma treatment are called plasma cleaned (P.C.).

\subsection{Fluid cell assembly and imaging}
An adhesive spacer (Grace Bio-Labs, 120~$\mu$m or 2600~$\mu$m height, 9~mm wells) was pressed onto the slide, and the well edges were sealed with a hydrophobic barrier pen (ImmEdge, Vector Laboratories). Immediately before assembly, 5~$\mu$L of 30\% H$_2$O$_2$ was added to 45~$\mu$L of particle suspension to give 3 wt./vol\% fuel. The well was filled (about 4~$\mu$L for the 120~$\mu$m cell and 110~$\mu$L for the 2600~$\mu$m cell) and sealed with a second glass slide by pressing once on the adhesive spacer. Recording started within 1~min of assembling the fluid cell. Videos (100~s at 10~fps, 20$\times$ objective, ThorLabs camera, 166.7~nm per pixel) were recorded alternately at the lowest and highest focal planes, giving five ceiling and five flooring videos per cell at different locations away from bubbles and cell edges. Each condition was measured in three separately prepared fluid cells.

\subsection{Particle tracking and speed}
Tracks were extracted with ImageJ/TrackMate using a custom Jython script that converted videos to 8-bit TIFF stacks, applied binary and Gaussian filters, detected particle centers and exported $(t, x, y)$ coordinates. Particles displaced by less than 20~pixels (3.3~$\mu$m) in 100~s were discarded as stuck. The mean-squared displacement of each track was fitted with
\begin{equation}
    \mathrm{MSD}(\tau) = 4D\tau + v^{2}\tau^{2},
    \label{eq:msd_short}
\end{equation}
the short-time limit ($\tau \ll \tau_R$) of the active Brownian expression
\begin{equation}
    \mathrm{MSD}(\tau) = 4D\tau
    + \frac{v^{2}\tau_{R}^{2}}{2}
      \left[\frac{2\tau}{\tau_{R}}
      + \exp\!\left(-\frac{2\tau}{\tau_{R}}\right) - 1\right],
    \label{eq:msd_full}
\end{equation}
where $D$ is the translational diffusion coefficient, $v$ the speed and $\tau_R$ the rotational relaxation time. For a sphere of radius $a = 1.5$~$\mu$m in water at 25~$^{\circ}$C, $D_R = k_BT/(8\pi\eta a^3) = 0.055$~s$^{-1}$ and $\tau_R = 1/D_R \approx 18$~s. Wall hindrance makes 18 s a lower bound. Fits were limited to $\tau \le 1$~s.

\subsection{Statistical analysis}
For each condition, speeds of ceiling and flooring particles from all three cells were pooled and the particle was treated as the unit of observation. Variances differed between conditions (Brown--Forsythe test, $W = 15.4$ for the pre-activation series and $W = 32.3$ for the batch comparison, both $p < 10^{-25}$), and group sizes were unequal, so Welch's one-way ANOVA (pre-activation: $F(10,\,363.6) = 173.4$; batches: $F(5,\,246.1) = 151.9$; both $p < 10^{-70}$) and the Games--Howell post hoc test\cite{games1976pairwise} were used. Results are shown as compact letter displays and $p$-value maps. Ceiling and flooring particles within a condition were compared with Welch's $t$-test, Holm-adjusted across the six batch conditions. A Games--Howell comparison of all twelve condition $\times$ plane groups gave the same conclusions. Eq.~\ref{eq:activation_dose} of the main text was fitted to condition means by weighted least squares (weights $1/\mathrm{SEM}^2$), parameter uncertainties were scaled by $\sqrt{\chi^2_\nu}$, and a 95\% interval for $E^*$ (5--28 wt./vol\%$\cdot$min) was obtained by resampling the three cells of each condition (2000 resamples). Significance was assessed at $\alpha = 0.05$. Analyses used Python with SciPy.

\subsection{X-ray photoelectron spectroscopy}
XPS was performed on untreated and plasma-cleaned glass and on platinum films on glass (7~nm) before and after immersion in 0.1\% H$_2$O$_2$ for 10~min, using a PHI 5000 VersaProbe II with monochromated Al~K$\alpha$ radiation (1486.6~eV), a 200~$\mu$m spot, a 45$^\circ$ take-off angle and dual-beam charge neutralization. Survey spectra (0--1100~eV, pass energy 93.9~eV, step 0.4~eV, 7--9 scans) were referenced to adventitious C~1s at 284.8~eV. Atomic percentages were obtained from peak areas with the manufacturer's sensitivity factors. Modified Auger parameters, which do not depend on charging, were calculated as the sum of the core-level binding energy and the kinetic energy of the corresponding KLL Auger line. Glass slides were cut to fit the sample stage.

\subsection{Conductivity}
Conductivity was measured with an Accumet AB200 meter calibrated with a 100~$\mu$S/cm standard. Five slides (75~$\times$~25~$\times$~1~mm), untreated or plasma cleaned, were immersed in 50~mL of ultra-pure water in a sealed polypropylene tube for 20~min at 22~$^\circ$C (0.40~mm$^{-1}$ glass area per volume). An identical tube without slides served as the control. Each condition was measured in triplicate.

\section{Estimate of the speed in plasma-cleaned cells}

This section explains how the speeds predicted in Figure~\ref{fig:xps-contact-angle-mechanism}c of the main text
were obtained. The estimate has three steps. First, we measured how much
plasma cleaning raises the conductivity of water in contact with glass. Second,
we scaled this increment to the treated area and fluid volume of each fluid
cell, allowing for the time that released ions need to diffuse across the
taller cell. Third, we converted the conductivity increment into a speed using
the relation between speed and conductivity measured by Brown and Poon for
Pt/PS Janus particles\cite{brown_ionic_2014}. As an independent check, we also
estimated the conductivity increment from the amount of fluoride that XPS
detected on plasma-cleaned glass.

\textbf{Step 1: conductivity added by plasma-cleaned glass.} Water in contact
with plasma-cleaned glass reached $2.27 \pm 0.08~\mu$S/cm and water in contact
with untreated glass $2.00 \pm 0.03~\mu$S/cm. The difference,
$\Delta\sigma_\mathrm{c} = 0.27 \pm 0.09~\mu$S/cm (uncertainties combined in
quadrature), is the conductivity added by plasma cleaning under the conditions
of this test. Each slide ($75 \times 25 \times 1$~mm) has a wetted area of
$2(75 \times 25) + 2(75 + 25)(1) = 3950$~mm$^2$, so five slides in 50~mL of water
give a glass area per fluid volume of
$(A/V)_\mathrm{c} = 19\,750/50\,000 = 0.40$~mm$^{-1}$.

\textbf{Background from the water and from untreated glass.} The water control
($0.83~\mu$S/cm) is higher than freshly dispensed 18.2~M$\Omega\cdot$cm water
($0.055~\mu$S/cm) because it absorbs CO$_2$ from the air. Carbonate equilibrium
(p$K_{a1} = 6.35$, p$K_{a2} = 10.33$) reproduces this value with about
12~$\mu$M total dissolved carbonate at pH 5.7. Untreated soda-lime glass
exchanges Na$^+$ from the glass network for H$_3$O$^+$ from the
water\cite{bunker1994molecular}. In CO$_2$-containing water, the hydroxide that
this exchange releases first neutralizes the carbonic acid, so the leachate
ends up mostly as sodium bicarbonate. The conductivity of $2.00~\mu$S/cm then
corresponds to about 15~$\mu$M Na$^+$ at pH $\approx$8.5, or
$\approx$39~$\mu$mol~m$^{-2}$ of glass, roughly the sodium contained in the outer
few nanometers of the glass. Brown and Poon found that adding NaOH did not
measurably slow Pt/PS particles\cite{brown_ionic_2014}, so we do not count this
alkaline leachate as a suppressing electrolyte. XPS shows less Na on
plasma-cleaned glass, which therefore probably leaches less Na than untreated
glass. The plasma-specific increment of $0.27~\mu$S/cm is then a lower bound on
the conductivity of the plasma-derived salt, and the predicted suppression is
conservative.

\textbf{Step 2: scaling to the fluid cells.} The increment was scaled to each
cell in proportion to its treated glass area per fluid volume,
\begin{equation}
  \Delta\sigma = \Delta\sigma_\mathrm{c}\,\frac{(A/V)_\mathrm{eff}}{(A/V)_\mathrm{c}},
\end{equation}
where $(A/V)_\mathrm{eff}$ is the effective treated area per volume of the cell.
For a cell with $N$ treated walls separated by a gap $h$, the treated area per
volume is $N/h$. This scaling assumes that the amount of salt released is
proportional to the treated area and that the contact time in the cells is
similar to that of the conductivity test (20~min).

Released ions spread across the gap by diffusion. With an ionic diffusion
coefficient $D_\mathrm{ion} \approx 1.5 \times 10^{-9}$~m$^2$/s, the slowest
diffusive mode across a gap $h$ relaxes on a time $h^2/(\pi^2 D_\mathrm{ion})$. In the 120~$\mu$m cell this time is $\approx 1$~s, so the released salt becomes
uniform across the gap within seconds and reaches both walls equally, as sketched
in Figure~\ref{fig:xps-contact-angle-mechanism}d of the main text; for this cell $(A/V)_\mathrm{eff} = N/h$. In the 2600~$\mu$m cell it is $\approx 8$~min, a third of the 25~min recording,
so unlike the situation in Figure~\ref{fig:xps-contact-angle-mechanism}d the concentration at the ceiling is not
uniform during much of the measurement. Let $\Gamma$ (mol~m$^{-2}$) be the amount of salt released per unit area of treated glass, assumed to dissolve at the moment the cell is filled. If this amount spread evenly across a gap $h$, the concentration would be $\Gamma/h$. For a wall at height $z_\mathrm{s}$, between two walls that neither absorb nor release further ions, the concentration at height $z$ and time $t$ is
\begin{equation}
  c(z,t) = \frac{\Gamma}{h}\left[1 + 2\sum_{n \ge 1}
  e^{-n^2\pi^2 D_\mathrm{ion} t/h^2}\cos\frac{n\pi z}{h}\cos\frac{n\pi z_\mathrm{s}}{h}\right].
\end{equation}
At the ceiling ($z = h$), a source at the floor ($z_\mathrm{s} = 0$) gives terms
of alternating sign, whereas a source at the ceiling itself ($z_\mathrm{s} = h$)
gives terms that are all positive. Averaging $c(h,t)$ over the recording, from 1
to 26~min after filling, gives $0.56\,\Gamma/h$ from a treated floor and
$1.63\,\Gamma/h$ from a treated ceiling. A treated floor alone therefore gives a
diffusion factor of 0.56, and two treated walls give $2.19\,\Gamma/h$, a factor
of 1.10 per treated wall. These factors multiply $N/h$ to give
$(A/V)_\mathrm{eff}$ (Table~\ref{tab:SI_scaling}). Starting the average at
filling instead gives factors of 0.52 and 1.25 and changes the predicted speeds
by less than 0.05~$\mu$m/s.

\begin{table}[!htbp]
\centering\small
\caption{Treated area per volume and conductivity increment in each cell. The diffusion factor is given per treated wall.}
\label{tab:SI_scaling}
\begin{tabular}{lcccc}
\toprule
Cell & $N/h$ (mm$^{-1}$) & Diffusion factor & $(A/V)_\mathrm{eff}$ (mm$^{-1}$) & $\Delta\sigma$ ($\mu$S/cm)\\
\midrule
120~$\mu$m, one wall   & 8.33  & 1    & 8.33  & 5.6\\
120~$\mu$m, two walls  & 16.67 & 1    & 16.67 & 11.3\\
2600~$\mu$m, one wall  & 0.385 & 0.56 & 0.22  & 0.15\\
2600~$\mu$m, two walls & 0.769 & 1.10 & 0.84  & 0.57\\
\bottomrule
\end{tabular}
\end{table}

\textbf{Step 3: from conductivity to speed.} In self-electrophoresis, the
reaction on the platinum cap drives an ionic current density $j$ through the
surrounding fluid. This current sets up an electric field
$E_\mathrm{f} = j/\sigma$, where $\sigma$ is the solution conductivity, and the
charged particle moves in this field at $v \approx \varepsilon\zeta E_\mathrm{f}/\eta$,
where $\varepsilon$ and $\eta$ are the permittivity and viscosity of the
solution and $\zeta$ is the zeta potential of the
particle\cite{paxton2006catalytically,moran2014role}. If $j$ and $\zeta$ do not
change, this contribution to the speed scales as $1/\sigma$. Brown and Poon
tested this prediction by adding NaCl or KBr to Pt/PS particles in 10\%
H$_2$O$_2$\cite{brown_ionic_2014}. The speed was linear in $1/\sigma$ but did not
extrapolate to zero: at high conductivity the particles kept a residual speed
$v_\mathrm{res} \approx 1.3~\mu$m/s, which the authors attributed tentatively to
neutral self-diffusiophoresis. Written relative to the untreated cell of the
same geometry, the speed in a plasma-cleaned cell is then
\begin{equation}
  v_\mathrm{pred} = v_\mathrm{res} + \left(v_\mathrm{u} - v_\mathrm{res}\right)\frac{\sigma_0}{\sigma_0 + \Delta\sigma},
  \label{eq:SI_vpred}
\end{equation}
where $v_\mathrm{u}$ is the measured speed in the untreated cell, $\sigma_0$ is the
conductivity of the fuel before any plasma-derived salt is added, and
$\Delta\sigma$ is the increment from Step 2. The first term is the part of the
speed that does not depend on conductivity, and the second is the
electrophoretic part, which falls as salt is added.

\textit{Fuel conductivity, $\sigma_0$.} We  bracketed $\sigma_0$ between 2 and 4~$\mu$S/cm. The upper bound is the 3.9~$\mu$S/cm
that Brown and Poon measured for 10\% H$_2$O$_2$\cite{brown_ionic_2014}; our 3\%
fuel is expected to be no more conductive. The lower bound is the conductivity
of water in contact with untreated glass ($2.00~\mu$S/cm), taken as the
background that the fuel acquires from glass under our conditions.

\textit{Residual speed, $v_\mathrm{res}$.} We did not measure $v_\mathrm{res}$ for
our particles and estimated it from the value of Brown and Poon in two steps.
First, we corrected for fuel concentration. Diffusiophoretic speed is
proportional to the rate of the surface reaction\cite{howse_self-motile_2007},
and Howse et al.\ found that this rate follows Michaelis--Menten kinetics,
$k = k_2 c/(c + k_2/k_1)$, where $c$ is the peroxide concentration and $k_1$ and
$k_2$ are the rate constants for formation and decomposition of a
Pt--H$_2$O$_2$ complex\cite{howse_self-motile_2007}. Their fitted constants give
$k_2/k_1 = 4.8\times10^{10}/4.4\times10^{11} = 0.109$. With $c$ expressed as a
volume fraction, this is a half-saturation concentration of about 11\%; this
reading reproduces the speed they measured at 10\% (about 3~$\mu$m/s) with their
fitted parameters, and it agrees with their observation that the speed was
still rising at 10\%. The reaction rate at 3\%
is therefore $[3/(3+11)]/[10/(10+11)] = 0.45$ times that at 10\%, and the
residual speed scales to $0.45 \times 1.3 = 0.59~\mu$m/s. Second, we considered
particle size. Our particles are 3~$\mu$m in diameter and those of Brown and Poon
were 2~$\mu$m; Ebbens et al.\ found that the speed of Pt/PS particles decreases
roughly in inverse proportion to particle radius in this size
range\cite{ebbens_size_2012}, which would lower the estimate to
$0.59 \times 2/3 \approx 0.4~\mu$m/s. We therefore used
$v_\mathrm{res} = 0.4$--$0.6~\mu$m/s, the same in both cell heights because it
reflects the particles and the fuel rather than the cell.

\textit{Worked example.} For the 120~$\mu$m one-wall cell
($v_\mathrm{u} = 2.65~\mu$m/s, $\Delta\sigma = 5.62~\mu$S/cm),
Eq.~\ref{eq:SI_vpred} gives
$0.4 + (2.65 - 0.4) \times 2/(2 + 5.62) = 0.99~\mu$m/s at the low ends of both
brackets and $0.6 + (2.65 - 0.6) \times 4/(4 + 5.62) = 1.45~\mu$m/s at the high
ends. The measured speed was $1.00 \pm 0.16~\mu$m/s. The ranges in
Table~\ref{tab:SI_velocity_prediction} and Figure~\ref{fig:xps-contact-angle-mechanism}c span both brackets. A single
measured $\sigma_0$ would narrow each range considerably.

\begin{figure}
    \centering
    \includegraphics[width=1\linewidth]{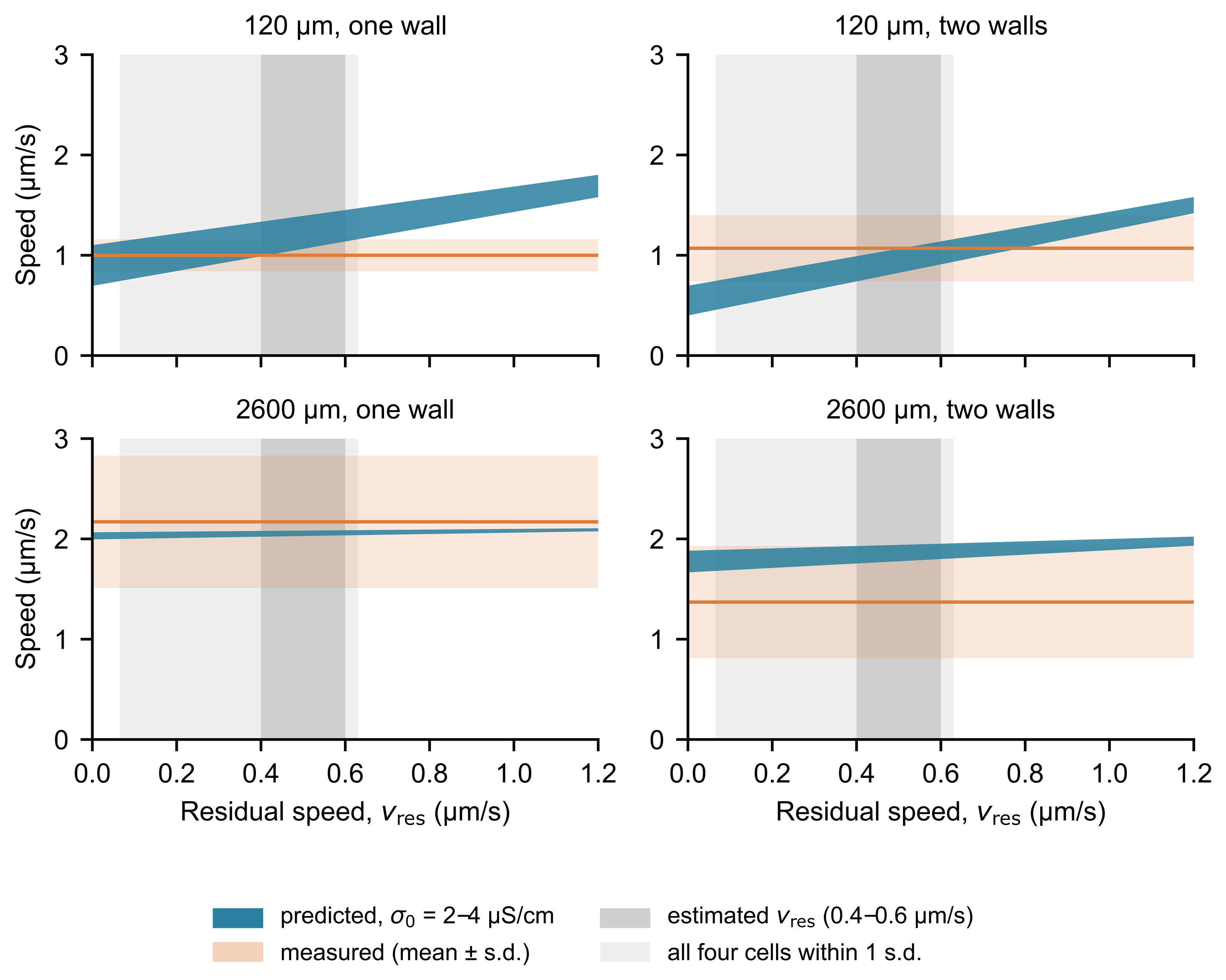}
    \caption{Conductivity-based speed predictions (teal bands; $\sigma_0 = 2$--$4~\mu$S/cm) against the residual speed $v_\mathrm{res}$ for each plasma-cleaned cell, with the measured mean $\pm$ s.d.\ of ceiling particles (orange). Dark grey: $v_\mathrm{res} = 0.4$--$0.6~\mu$m/s, estimated from the residual reported by Brown and Poon scaled to 3\% H$_2$O$_2$ and to the size of our particles. Light grey: range of $v_\mathrm{res}$ (0.07--0.63~$\mu$m/s) for which all four predictions lie within one standard deviation of the measurements.}
    \label{fig:SI_residual}
\end{figure}

\textit{Sensitivity to the residual speed.} Setting $v_\mathrm{res} = 0$, as in
a purely electrophoretic model, lowers every prediction. The 2600~$\mu$m cells
change by less than 0.05~$\mu$m/s because $\Delta\sigma$ is much smaller than
$\sigma_0$ there, but the prediction for the 120~$\mu$m two-wall cell falls to
0.40--0.70~$\mu$m/s, below the measured $1.07 \pm 0.33~\mu$m/s.
Figure~\ref{fig:SI_residual} shows the conductivity-based predictions as a
function of $v_\mathrm{res}$. All four predictions lie within one standard
deviation of the measurements for $v_\mathrm{res}$ between 0.07 and
0.63~$\mu$m/s. A residual as large as 1~$\mu$m/s would overpredict the speed in
the 120~$\mu$m one-wall cell (1.43--1.69 versus $1.00 \pm 0.16~\mu$m/s).

\textbf{Independent check from XPS.} The conductivity test shows how much
plasma-cleaned glass adds to the conductivity of water, but not which ions are
responsible. As a check, we asked whether the fluoride that XPS detected on
plasma-cleaned glass would, if it all dissolved, add about the measured
$0.27~\mu$S/cm. The calculation has three parts.

\textit{(i) Amount of fluorine on the surface.} XPS reports composition in atomic
percent, the fraction of all detected atoms that are fluorine. Plasma cleaning
raised this fraction from 0.5 to 3.2~at.\%, so the fluorine added by plasma
cleaning is $x_\mathrm{F} = 2.7$~at.\% $= 0.027$. To turn this fraction into an
amount per unit area of glass, $\Gamma_\mathrm{F}$, we need to know how deep the
fluorine sits. XPS collects signal from a layer a few nanometers thick and
weights atoms near the surface more strongly than deeper ones: the signal from
atoms at depth $z$ falls as $\exp(-z/d)$, where
$d = \lambda_\mathrm{att}\cos\theta = 1.84$~nm is the effective probing depth,
$\lambda_\mathrm{att} = 2.6$~nm is the attenuation length of F~1s photoelectrons
in silica, and $\theta = 45^\circ$ is the take-off angle. About 95\% of the
signal comes from the top $3d = 5.5$~nm. Because XPS does not show how the
fluorine is distributed within this depth, we calculated $\Gamma_\mathrm{F}$ for
two limiting cases that bracket any real distribution. Both use the atomic
density of glass, $n_\mathrm{g} = 7 \times 10^{22}$~atoms~cm$^{-3}$.

\emph{Limit 1: fluorine spread evenly through the probed layer.} If fluorine
makes up a fraction $x_\mathrm{layer}$ of the atoms throughout the top 5.5~nm,
this layer supplies 95\% of the signal, so $x_\mathrm{F} = 0.95\,x_\mathrm{layer}$
and $x_\mathrm{layer} = 0.027/0.95 = 0.028$. The fluorine per unit area is then
$\Gamma_\mathrm{F} = x_\mathrm{layer}\,n_\mathrm{g}\,(3d)
= 0.028 \times (7 \times 10^{22}~\mathrm{cm^{-3}}) \times (5.5 \times 10^{-7}~\mathrm{cm})
= 1.1 \times 10^{15}$~atoms~cm$^{-2}$, or 18~$\mu$mol~m$^{-2}$.

\emph{Limit 2: fluorine only in the outermost atomic layer} ($t = 0.25$~nm).
This thin layer supplies only $1 - \exp(-t/d) = 12.7$\% of the signal, so to
give the measured 2.7~at.\% it must be rich in fluorine,
$x_\mathrm{layer} = 0.027/0.127 = 0.21$. Then
$\Gamma_\mathrm{F} = 0.21 \times (7 \times 10^{22}~\mathrm{cm^{-3}}) \times (2.5 \times 10^{-8}~\mathrm{cm})
= 3.7 \times 10^{14}$~atoms~cm$^{-2}$, or 6.2~$\mu$mol~m$^{-2}$.

The plasma-cleaned glass therefore carries 6--18~$\mu$mol of fluorine per square
meter.

\textit{(ii) Concentration in solution.} If all of this fluoride dissolved into
the water of the conductivity test, where the glass area per volume was
$(A/V)_\mathrm{c} = 0.40$~mm$^{-1} = 400$~m$^{-1}$, its concentration would be
$[\mathrm{F}^-] = \Gamma_\mathrm{F}\,(A/V)_\mathrm{c}
= (6.2\mbox{--}18~\mu\mathrm{mol~m^{-2}}) \times (400~\mathrm{m^{-1}})
= 2.5$--$7.3~\mu$M. As CaF$_2$, which releases two F$^-$ for each Ca$^{2+}$,
this corresponds to 1.2--3.6~$\mu$M CaF$_2$.

\textit{(iii) Conductivity added.} The conductivity of a dilute salt solution is
its molar conductivity times its concentration. For fully dissociated CaF$_2$
the molar conductivity is the sum of the tabulated limiting molar conductivities
of its ions, $\Lambda_{\mathrm{CaF}_2} = \lambda^\circ_{\mathrm{Ca}^{2+}} + 2\lambda^\circ_{\mathrm{F}^-}
= 119.0 + 2(55.4) = 229.8$~S~cm$^2$~mol$^{-1}$. The added conductivity is then
$\Delta\sigma = \Lambda_{\mathrm{CaF}_2}\,[\mathrm{CaF}_2]
= (229.8~\mathrm{S~cm^2~mol^{-1}}) \times (1.2\mbox{--}3.6 \times 10^{-9}~\mathrm{mol~cm^{-3}})
= 0.28$--$0.84~\mu$S/cm.

The measured increment, $0.27~\mu$S/cm, sits at the lower end of this range.
Working backwards, it corresponds to $1.2~\mu$M CaF$_2$ (2.35~$\mu$M F$^-$), or
5.9~$\mu$mol of fluoride released per square meter of glass, close to the
thin-layer limit. The fluoride on plasma-cleaned glass is therefore enough to
account for the measured increment.

For the XPS-based speeds in Table~\ref{tab:SI_velocity_prediction}, the same
calculation was repeated for each fluid cell with its own treated area per
volume, $[\mathrm{CaF}_2] = \Gamma_\mathrm{F}(A/V)_\mathrm{eff}/2$, and the
resulting $\Delta\sigma$ was used in Eq.~\ref{eq:SI_vpred}. The ranges combine
the two inventory limits with the brackets on $\sigma_0$ and $v_\mathrm{res}$.
The highest concentration, 0.15~mM CaF$_2$ in the 120~$\mu$m two-wall cell, is
below the solubility of CaF$_2$ ($\approx$0.21~mM, from
$K_\mathrm{sp} \approx 3.9 \times 10^{-11}$), so complete dissolution is
possible in every cell.

\begin{table}[!htbp]
\centering\small
\caption{Predicted and measured speeds of ceiling particles in plasma-cleaned cells}
\label{tab:SI_velocity_prediction}
\begin{tabular}{lccccc}
\toprule
Cell & $v_\mathrm{u}$ & $\Delta\sigma$, conductivity & $v_\mathrm{pred}$, conductivity & $v_\mathrm{pred}$, XPS & $v_\mathrm{meas}$\\
     & ($\mu$m/s) & ($\mu$S/cm) & ($\mu$m/s) & ($\mu$m/s) & ($\mu$m/s)\\
\midrule
120~$\mu$m, one wall   & $2.65\pm0.62$ & 5.6  & 0.99--1.45 & 0.63--1.43 & $1.00\pm0.16$\\
120~$\mu$m, two walls  & $2.65\pm0.62$ & 11.3 & 0.74--1.14 & 0.52--1.12 & $1.07\pm0.33$\\
2600~$\mu$m, one wall  & $2.14\pm0.53$ & 0.15 & 2.02--2.09 & 1.82--2.08 & $2.17\pm0.66$\\
2600~$\mu$m, two walls & $2.14\pm0.53$ & 0.57 & 1.75--1.95 & 1.32--1.94 & $1.37\pm0.56$\\
\bottomrule
\end{tabular}

\vspace{1ex}\footnotesize
Measured values are mean $\pm$ s.d.\ of ceiling particles. Predicted ranges span
$\sigma_0 = 2$--$4~\mu$S/cm and $v_\mathrm{res} = 0.4$--$0.6~\mu$m/s; XPS-based
ranges also include the two limits of the fluoride inventory.
\end{table}

\section{Supplementary figures and tables}

\begin{figure}
    \centering
    \includegraphics[width=1\linewidth]{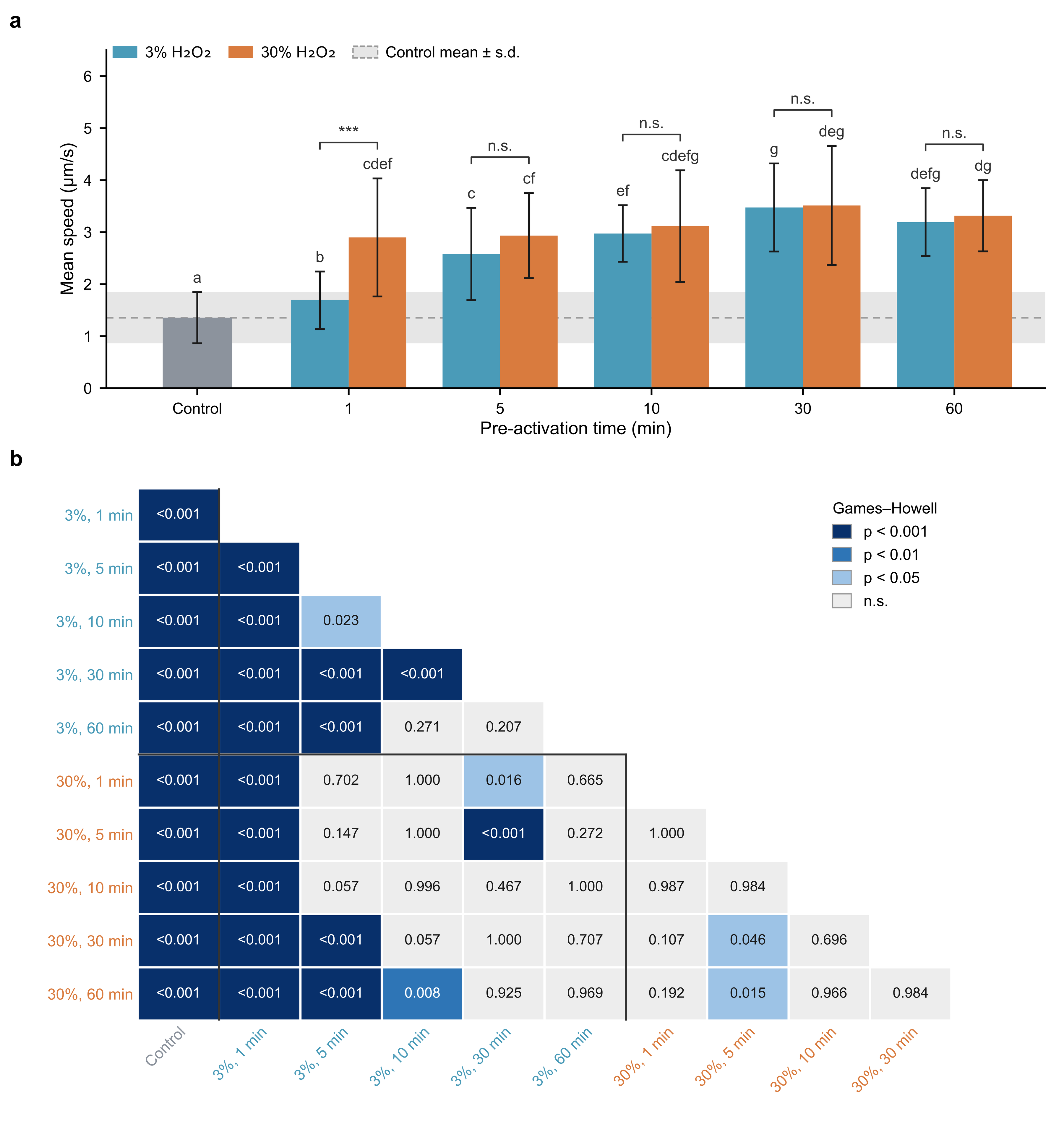}
    \caption{Statistical comparison of pre-activation conditions. (a) Mean speed (error bars, s.d.) of ceiling and flooring particles pooled from three cells ($n = 55$--$194$). The shaded band is the non-activated control, mean $\pm$ s.d. Conditions sharing a letter do not differ; brackets compare 3\% and 30\% activation at equal time (*** $p < 0.001$; n.s., not significant). (b) Games--Howell adjusted $p$-values for all 55 pairwise comparisons. $\alpha = 0.05$.}
    \label{fig:si-activation-time}
\end{figure}

\begin{figure}
    \centering
    \includegraphics[width=1\linewidth]{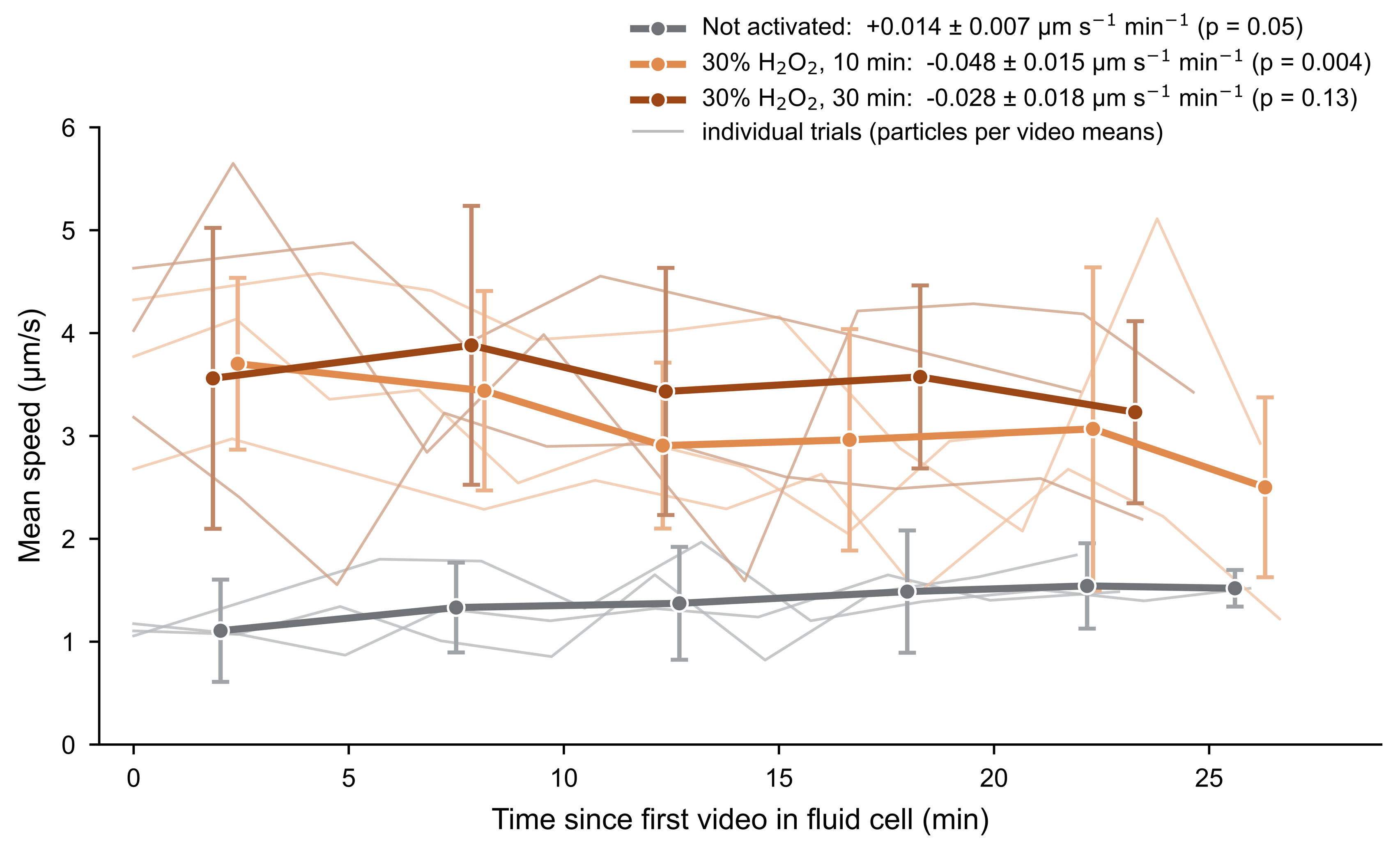}
    \caption{Speed against time after assembly of the fluid cell (120~$\mu$m spacer, untreated glass, 3 wt./vol\% H$_2$O$_2$) for non-activated particles (grey; 3 cells, 30 videos, 194 tracks) and particles pre-activated in 30\% H$_2$O$_2$ for 10~min (light orange; 3 cells, 30 videos, 63 tracks) or 30~min (dark orange; 3 cells, 25 videos, 55 tracks). Bold lines are the mean speed of all tracks in 5~min bins (ceiling and flooring pooled; error bars, s.d.; bins with fewer than 3 tracks omitted). Faint lines are video means for individual cells. Time zero is the start of the first video in each cell. Legend values are slopes ($\pm$ standard error) of weighted linear fits of video means against time, with a separate intercept for each cell and plane; $p$-values test whether each slope differs from zero.}
    \label{fig:si-with-time-control-activated}
\end{figure}

\begin{figure}
    \centering
    \includegraphics[width=1\linewidth]{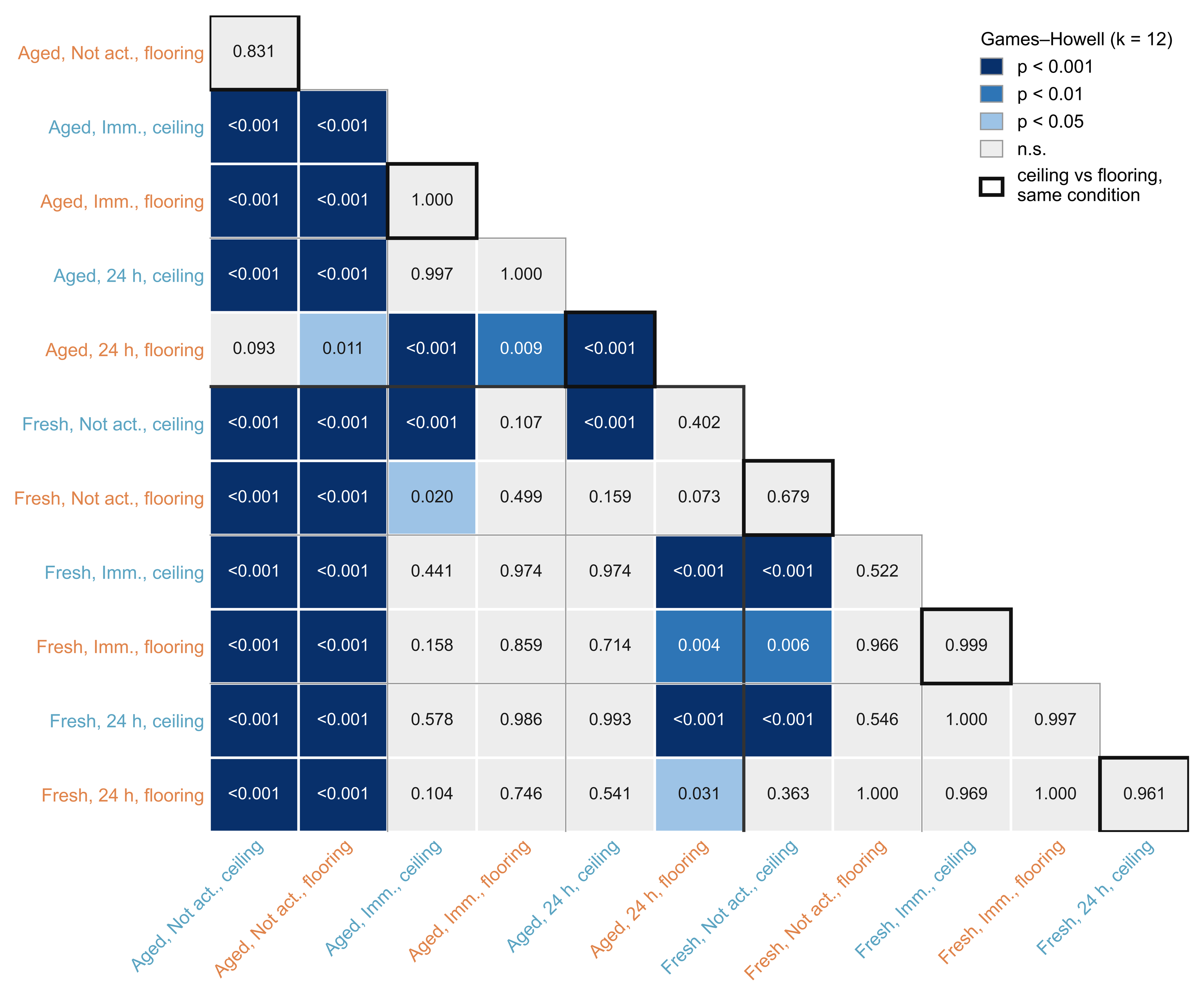}
    \caption{Games--Howell adjusted $p$-values for all 66 pairwise comparisons among the twelve condition $\times$ imaging-plane groups of the aged and fresh batches (not pre-activated, pre-activated and recorded immediately, or pre-activated and stored for 24~h; three cells pooled per group). Outlined cells compare ceiling and flooring within a condition. Thick lines separate the batches; thin lines separate conditions. $\alpha = 0.05$.}
    \label{fig:aged-fresh-heatmap-p-values}
\end{figure}

\begin{table}[!htbp]
\centering\small
\caption{Tracked particles and mean speed for aged and fresh batches}
\label{tab:SI_aged_fresh_desc}
\begin{tabular}{lrcrcrc}
\toprule
Condition & $n$ & $\bar{v}$ ($\mu$m/s) & $n_C$ & $\bar{v}_C$ ($\mu$m/s) & $n_F$ & $\bar{v}_F$ ($\mu$m/s)\\
\midrule
Aged, not pre-activated & 194 & $1.36\pm0.49$ & 94 & $1.42\pm0.48$ & 100 & $1.30\pm0.50$ \\
Aged, immediate & 63 & $3.12\pm1.07$ & 35 & $3.15\pm0.92$ & 28 & $3.07\pm1.26$ \\
Aged, 24 h storage & 100 & $2.54\pm1.11$ & 60 & $2.95\pm1.02$ & 40 & $1.92\pm0.95$ \\
Fresh, not pre-activated & 181 & $2.34\pm0.44$ & 139 & $2.30\pm0.39$ & 42 & $2.49\pm0.57$ \\
Fresh, immediate & 96 & $2.71\pm0.53$ & 58 & $2.75\pm0.57$ & 38 & $2.65\pm0.47$ \\
Fresh, 24 h storage & 81 & $2.68\pm0.62$ & 46 & $2.77\pm0.61$ & 35 & $2.57\pm0.62$ \\
\bottomrule
\end{tabular}

\vspace{1ex}\footnotesize
Mean $\pm$ s.d.; $n$, all particles; subscripts C and F, ceiling and flooring.
\end{table}

\begin{figure}
    \centering
    \includegraphics[width=1\linewidth]{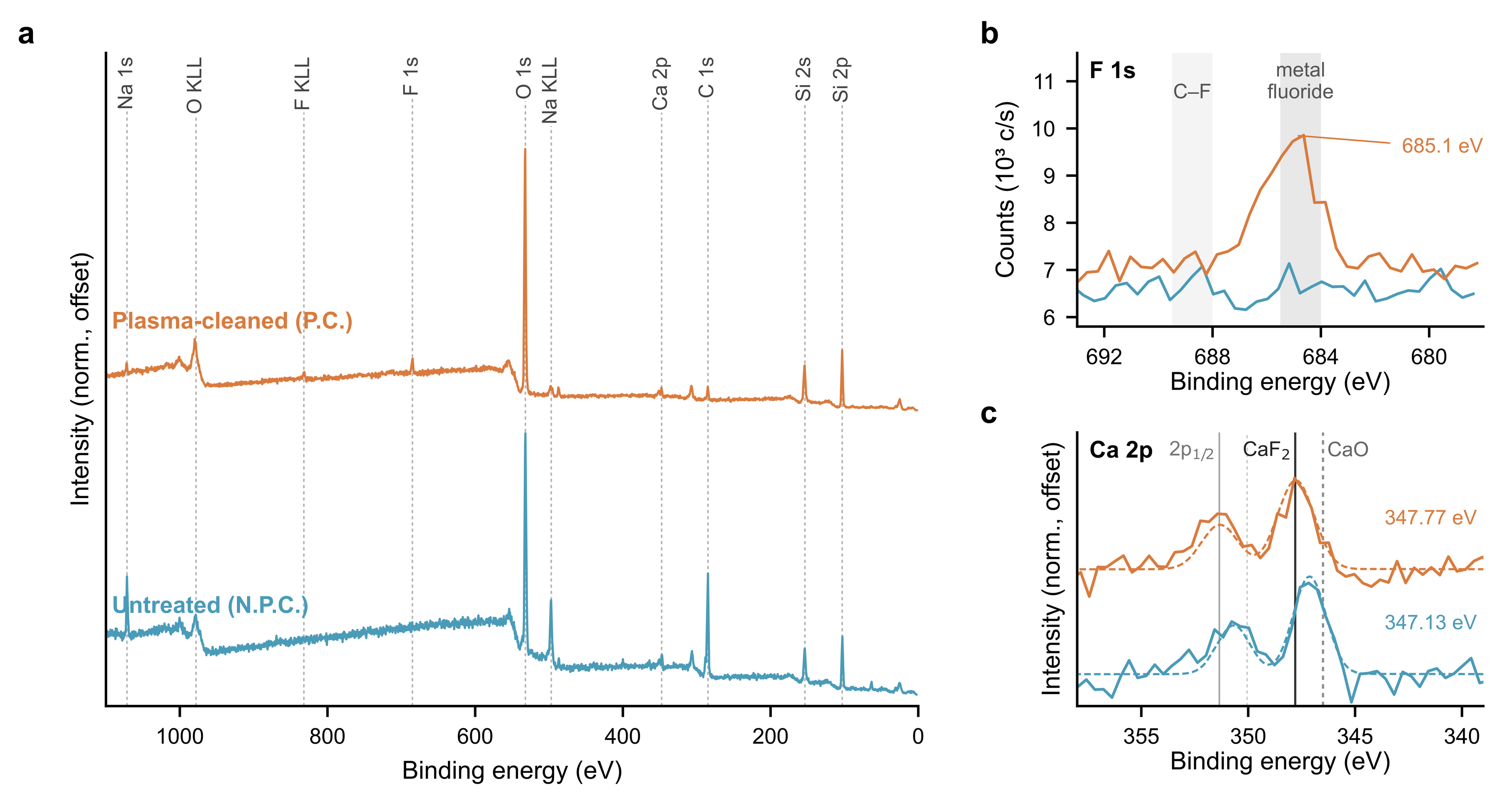}
    \caption{XPS of untreated (N.P.C.) and plasma-cleaned (P.C.) glass, referenced
to adventitious C~1s at 284.8~eV. (a) Survey spectra normalized to the O~1s
maximum and offset. (b) F~1s region. Shaded bands mark the binding-energy
ranges of metal fluorides and of C--F bonds. (c) Ca~2p region with
spin--orbit doublet fits (dashed; 2p$_{3/2}$--2p$_{1/2}$ splitting 3.55~eV);
vertical lines mark the Ca~2p$_{3/2}$ and 2p$_{1/2}$ positions of CaF$_2$
(solid) and CaO (dashed). The fitted Ca~2p$_{3/2}$ position shifts from 347.13
to 347.77~eV after plasma cleaning.}
\label{fig:SI_glass_XPS}
\end{figure}

\begin{figure}
    \centering
    \includegraphics[width=1\linewidth]{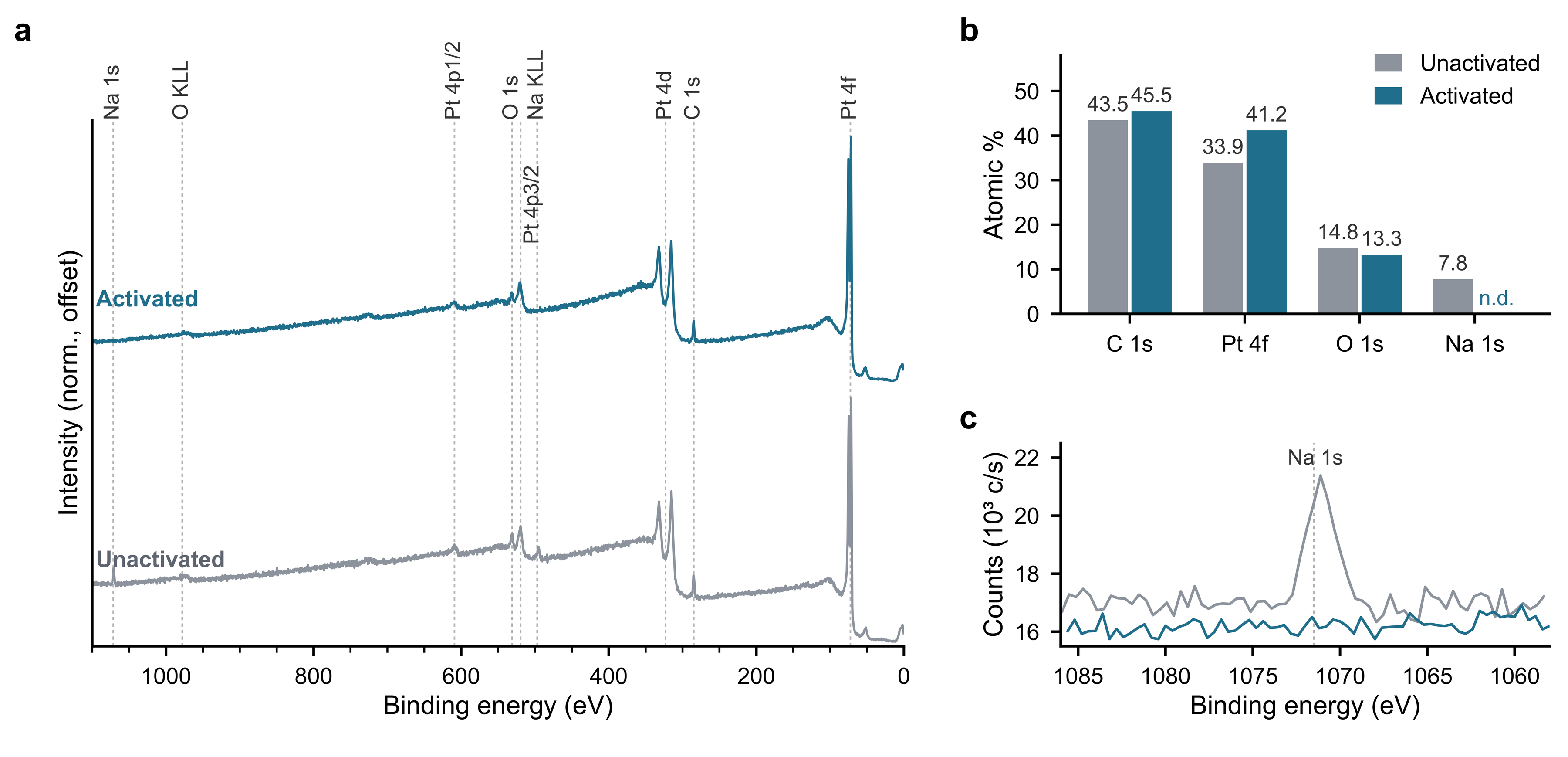}
    \caption{XPS of the platinum film before and after exposure to 0.1\% H$_2$O$_2$ for 10~min. (a) Survey spectra referenced to C~1s at 284.8~eV, normalized to the Pt~4f maximum and offset. (b) Surface composition; n.d., not detected. The O~1s window overlaps Pt~4p$_{3/2}$, so O contents are approximate. (c) Na~1s region, showing loss of the Na signal after exposure.}
    \label{fig:SI_Pt_XPS}
\end{figure}

\end{document}